\documentclass[fleqn]{cas-sc}

\usepackage{amssymb}
\usepackage{amsmath}
\usepackage{amsfonts}
\usepackage{lineno}
\modulolinenumbers[5]
\usepackage[numbers,sort]{natbib}
\usepackage{textgreek}
\usepackage{graphicx}
\usepackage{grffile}
\DeclareGraphicsExtensions{.pdf,.jpeg,.jpg,.png}
\usepackage{xcolor}
\usepackage{array}
\usepackage{multirow}
\usepackage{tabulary}
\usepackage{booktabs}
\usepackage{setspace}
\usepackage{algorithm}
\usepackage{algorithmicx}
\usepackage{algpseudocode}

\usepackage{tikz}
\usetikzlibrary{arrows.meta}
\usetikzlibrary{decorations.pathmorphing}
\usepackage[utf8]{inputenc}
\usepackage{ifthen}
\usepackage{pgfplots}
\DeclareUnicodeCharacter{2212}{−}
\usepgfplotslibrary{groupplots,dateplot}
\usetikzlibrary{patterns,shapes.arrows}
\pgfplotsset{compat=1.18}

\newcommand{\isot}[2]{\(^{\mathrm{#2}}\){#1}}

\DeclareMathOperator*{\argmax}{arg\, max}
\DeclareMathOperator*{\argmin}{arg\, min}

\usepackage{soul}
\usepackage{etoolbox}
\makeatletter
\patchcmd{\SOUL@ulunderline}{\dimen@}{\SOUL@dimen}{}{}
\patchcmd{\SOUL@ulunderline}{\dimen@}{\SOUL@dimen}{}{}
\patchcmd{\SOUL@ulunderline}{\dimen@}{\SOUL@dimen}{}{}
\newdimen\SOUL@dimen
\makeatother

\usepackage{url}
\usepackage{hyperref}
\usepackage[plain]{fancyref}

\begin{document}
\let\WriteBookmarks\relax
\def\floatpagepagefraction{1}
\def\textpagefraction{.001}

\shorttitle{Full Spectrum Modeling of Precipitation Transients}

\shortauthors{M.\,S.~Bandstra, et al.}

\title[mode=title]{Full Spectrum Modeling of \textit{In Situ} Gamma-ray Detector Measurements with a Focus on Precipitation-Induced Transients}


\author[1]{M. S. Bandstra}[orcid=0000-0002-6403-7895]
\cormark[1]
\ead{msbandstra@lbl.gov}
\credit{Conceptualization, Formal analysis, Investigation, Methodology, Software, Validation, Visualization, Writing -- original draft}

\author[2,3]{J. M. Ghawaly}[orcid=0000-0001-8826-8500]
\credit{Conceptualization, Data curation, Methodology, Writing -- original draft}

\author[2]{D. E. Peplow}[orcid=0000-0002-9208-1914]
\credit{Methodology}

\author[2]{D. E. Archer}[orcid=0000-0003-3397-6109]
\credit{Data curation, Funding acquisition, Supervision, Writing -- original draft}

\author[1]{B. J. Quiter}[orcid=0000-0001-7001-7455]
\credit{Funding acquisition, Supervision}

\author[1]{T. H. Y. Joshi}[orcid=0000-0003-0846-7871]
\credit{Conceptualization, Funding acquisition, Methodology, Supervision}


\author[2]{A. D. Nicholson}[orcid=0000-0001-5303-5424]
\credit{Data curation}

\author[2]{M. J. Willis}[orcid=0000-0003-3382-1174]
\credit{Data curation, Methodology}

\author[2]{I. Garishvili}[orcid=0000-0001-9673-3310]
\credit{Data curation}

\author[4]{A. J. Rowe}
\credit{Data curation}

\author[2]{B. R. Longmire}[orcid=0000-0001-9714-6361]
\credit{Data curation}


\author[2]{J. T. Nattress}[orcid=0000-0001-5449-2520]
\credit{Methodology, Validation}

\affiliation[1]{%
    organization={Nuclear Science Division, Lawrence Berkeley National Laboratory},
    addressline={1 Cyclotron Road},
    city={Berkeley},
    state={CA},
    postcode={94720},
    country={USA},
}

\affiliation[2]{%
    organization={Physics Division, Oak Ridge National Laboratory},
    addressline={P.O. Box 2008},
    city={Oak Ridge},
    state={TN},
    postcode={37831},
    country={USA},
}

\affiliation[3]{
    organization={Computer Science and Engineering Division, Louisiana State University},
    addressline={Patrick F. Taylor Hall},
    city={Baton Rouge},
    state={LA},
    postcode={70803},
    country={USA},
}

\affiliation[4]{%
    organization={Cadre5}
}

\begin{abstract}
    Gamma-ray detectors that are deployed outdoors experience increased event rates during precipitation due to the attendant increase in Rn-222 progeny at ground level.
    The increased radiation due to these decay products (Pb-214 and Bi-214) has been studied for many decades in applications such as atmospheric science and radiation protection.
    For those applications radon progeny signatures are the signal of interest, while in the fields of radiological and nuclear security and aerial radiological mapping they are a nuisance.
    When searching for radiological contamination or missing sources, an analyst must take precipitation into account to reduce false alarms, in addition to accounting for static background signatures.
    To train advanced search algorithms, an effort has been underway to generate synthetic gamma-ray event data that represent a realistic urban area, including occasional rain events to add to the realism.
    This manuscript describes an effort to analyze and model gamma-ray spectra measured during rainfall by a NaI(Tl) detector located outdoors in order to derive accurate source terms for Pb-214 and Bi-214 at a high frequency (less than 1 minute).
    All known sources of background were quantitatively modeled across the full gamma-ray spectrum, so that the Pb-214 and Bi-214 activity concentrations on the ground could be inferred from a linear model fit to each spectrum.
    A physically motivated model was applied to the data to further smooth the fits, which had the benefit of yielding information about the concentrations of the progeny in rainwater and their apparent age, making this the first time full-spectrum modeling has been used for continuous measurements of radon progeny.
    Full-spectrum modeling's ability to leverage more statistics allows for measurements at a rate of more than once per minute, rather than the more typical 10- or 15-minute measurement cycle, and therefore this approach could lead to studies of radon progeny on shorter timescales than previously possible.
\end{abstract}


\maketitle



\section{Introduction}\label{sec:introduction}
After the first discoveries of the natural radioactivity of uranium and thorium compounds~\cite{becquerel_radiations_1896,curie_rayons_1898,schmidt_uber_1898}, some noticed that the compounds exhibited radioactive ``emanations,'' which are now known as radon gas~\cite{rutherford_thorium_1899,dorn_uber_1901}.
In 1901, these same emanations were found to occur not just in the laboratory with specific compounds but also ubiquitously in the outside air~\cite{elster_uber_1901,geitel_uber_1901}.
Shortly thereafter, in 1902, C.\,T.\,R.~Wilson discovered that rain is radioactive~\cite{wilson_radio-active_1902,wilson_further_1902}.
It was soon conjectured that the radioactivity of precipitation was due to the emanations of radium~\cite{bumstead_atmospheric_1904}, and this was confirmed in subsequent years~\cite{gockel_beobachtungen_1908}.
A review of the early history of observations of the radioactivity of precipitation can be found in ref.~\cite{damon_natural_1954}.

The emanation of radium is now known as the noble gas \isot{Rn}{222} and its decay products, which come from the \isot{U}{238} decay chain.
The relevant decay products or progeny of \isot{Rn}{222}, which has a half-life of 3.821~days, are \isot{Po}{218}, \isot{Pb}{214}, and \isot{Bi}{214} (half-lives 3.098\,min, 27.06\,min, and 19.90\,min, respectively), the first of which is an alpha emitter and the latter two are beta and gamma-ray emitters~\cite{nndc_nudat}.

The emanation of thorium is now known to be \isot{Rn}{220} and its decay products, which come from the \isot{Th}{232} decay chain~\cite{rutherford_radioactivity_1900}.
It is also present in the environment in appreciable quantities, however its contribution to the radioactivity of rain can be safely neglected for our purposes.
\isot{Rn}{220}, also called thoron, has a 55.6-second half-life that severely limits its exhalation rate from the ground to roughly 1/80 of the rate of \isot{Rn}{222}, and restricts its vertical distribution to primarily the lowest few meters above ground~\cite{jacobi_vertical_1963}.
Its only decay product of significance is \isot{Pb}{212} (half-life: 10.64\,h), which is often found at higher concentrations than \isot{Rn}{220} in the lower atmosphere due to its longer half-life~\cite{jacobi_vertical_1963}.
However, its concentration at cloud altitudes is likewise negligible, leading to rainwater concentrations more than two orders of magnitude lower than the progeny of \isot{Rn}{222}~\cite{rangarajan_observations_1985}.
Although they are only a negligible contribution to the outdoor gamma-ray background and to the radioactivity of rain, \isot{Rn}{220} and its decay products can still be a significant health hazard due to their presence in indoor air~\cite{porstendorfer_properties_1994}.

A third isotope of radon also occurs in nature.
Actinon (\isot{Rn}{219}), which comes from the \isot{U}{235} decay chain and has a half-life of 3.96\,s, is even further suppressed due to its short half-life and the low natural abundance of \isot{U}{235}, so it will also be neglected in this work.

Because the half-lives of \isot{Rn}{222} progeny range from a few minutes to a few tens of minutes, which is the same order as the time it takes a raindrop to condense and fall from a cloud to the earth, the activity concentration in raindrops encodes information about the atmospheric conditions under which the precipitation is occurring~\cite{greenfield_determination_2008}.
Over the years, detailed models of the evolution of radon progeny in precipitation have been developed~\cite{damon_natural_1954,jacobi_anlagerung_1961,takeuchi_rainout-washout_1982,minato_estimate_1983,horng_rainout_2003,minato_simple_2007}, as well as various kinds of continuous measurement systems, both ones that repeatedly collect and count samples of rainwater~\cite{nishikawa_automatic_1986,paatero_wet_2000,cortes_automated_2001,takeyasu_concentrations_2006,liu_characteristics_2014} and systems that measure the gamma-ray events \textit{in situ} as they come from the environment around the system~\cite{finck_situ_1980,minato_analysis_1980,katase_variation_1982,hayakawa_radon-concentration--cloud_1985,horng_situ_2004}.
For instance, studies have been able to show that radon progeny levels vary seasonally~\cite{minato_estimate_1983,hayakawa_radon-concentration--cloud_1985,fujinami_observational_1996}, and radon progeny concentrations tend to scale inversely proportional to rainfall rate~\cite{damon_natural_1954,bleichrodt_dependence_1959,bhandari_study_1963,fujinami_observational_1996}.
The ratio of \isot{Bi}{214} to \isot{Pb}{214} has also been used to estimate the radiometric age of the rain~\cite{horng_rainout_2003,greenfield_determination_2008,moriizumi_214bi214pb_2015}.

In addition, studies have shown that ``rainout'' --- i.e., progeny already having been incorporated into raindrops before they fall from the cloud --- is a much more important physical process than ``washout'' --- i.e., progeny being collected by raindrops as they fall --- since radon progeny concentrations in the air at ground level tend not to change much during rain~\cite{fujinami_observational_1996}.
Therefore, the radon progeny in rain are primarily coming from the decay of radon at altitude in the cloud, which has been confirmed in studies showing that higher radon progeny levels are seen in rain when the air mass comes from over land rather than over the sea~\cite{paatero_wet_2000,yamazawa_wet_2008}, and correlations have even been seen with the integrated radon emanation rate of the land that the air mass has passed over~\cite{mercier_increased_2009}.

Although the first measurements of radon progeny in rain were done with electroscopes~\cite{wilson_radio-active_1902,priebsch_uber_1932} and Geiger-M\"{u}ller tubes~\cite{damon_natural_1954,bleichrodt_dependence_1959}, detectors offering spectroscopic information have proved useful in separating the signatures of \isot{Pb}{214} from \isot{Bi}{214}, especially in being able to separate the mostly lower-energy photons of \isot{Pb}{214} from the higher-energy photons of \isot{Bi}{214}, or, for greater accuracy, to even measure the net counts in the most prominent photopeaks of each isotope (352 and 609~keV, respectively).
The earliest spectroscopic detectors used were scintillators, most commonly NaI(Tl)~\cite{bhandari_study_1963,fujinami_influence_1985,hayakawa_radon-concentration--cloud_1985,nishikawa_automatic_1986,paatero_wet_2000,liu_characteristics_2014,patiris_atmospheric_2023}, while in later years Ge(Li)~\cite{finck_situ_1980,takeuchi_rainout-washout_1982} and more recently high-purity germanium (HPGe)~\cite{cortes_automated_2001,horng_situ_2004,takeyasu_concentrations_2006,greenfield_determination_2008} have also been used.

Apart from its use in atmospheric studies, the fact that precipitation is radioactive is generally either used as a tool or seen as a nuisance.
For example, in agricultural applications such as proximal gamma-ray spectroscopy (PGRS), outdoor gamma-ray spectrometers are used to measure soil properties~\cite{mahmood_proximal_2013} and must contend with rain-induced signatures.
Full-spectrum models of gamma-ray background have been used in PGRS analysis to determine soil properties~\cite{baldoncini_investigating_2018}, while net peak counts have been used to distinguish irrigation from rainfall by comparing the relative changes to the \isot{Pb}{214} photopeak at 352\,keV and the \isot{K}{40} photopeak at 1460\,keV~\cite{serafini_proximal_2021}.
PGRS has also been used to perform detailed long-term studies of radon progeny in rainfall~\cite{bottardi_rain_2020}.
Aerial radiological surveys similarly must contend with radon progeny from rain~\cite{amestoy_effects_2021}, and are advised to wait at least three hours after rainfall before performing measurements~\cite{minty_fundamentals_1997}.

In nuclear and radiological security applications, the fact that rain is radioactive complicates the work of researchers.
For example, radiation portal monitors, which are used to screen vehicles for radioactivity in their cargo and are generally located outdoors, can suffer from large changes in background when it rains, reducing their sensitivity to anomalies~\cite{livesay_rain-induced_2014}.
Other stationary monitoring detectors~\cite{hoteling_analysis_2021,bandstra_background_2023} as well as mobile search systems~\cite{aucott_routine_2013,mitchell_gamma-ray_2015} similarly experience significant background shifts during rain that can affect their detection performance.
Because of its impact on this area of research, several efforts to model and even predict temporally varying rain signatures have been undertaken in this context~\cite{liu_spatial-temporal_2018,liu_prediction_2019,livesay_rain-induced_2014,bandstra_background_2023}.

Because of the importance of radioactive source search to the security mission, a significant amount of research has been done to study the varying natural backgrounds encountered by static and mobile detector systems~\cite{sandness_accurate_2009,archer_modeling_2017,nicholson_characterization_2018_minos1,salathe_determining_2021}.
A major recent effort has been the generation of synthetic gamma-ray event data to mimic a realistic urban environment, with the intention being to replicate in Monte Carlo simulation all of the dominant sources of natural radioactivity that a mobile system may encounter, including their temporal variability~\cite{ghawaly_data_2020,nicholson_generation_2020,peplow_monte_2021,radai_datasets_webpage}.
The effort described here was undertaken to add realistic rainfall effects to the synthetic urban background dataset to help make future search algorithms more robust to the effects of rain.
The basic approach was to take data using a stationary, outdoor gamma-ray detector and use it to estimate the time-varying source terms of \isot{Pb}{214} and \isot{Bi}{214} on the ground in units of Bq/m\(^2\).
These source terms were then used to simulate sources on the ground in the synthetic urban scene.
Because backgrounds in mobile detector systems can vary on the order of \( \sim \)1\,s, it was important to model these rain events on as short a timescale as practical, which we determined to be approximately 1~minute.

This work describes the effort to ``invert'' the time-varying measured spectra from a stationary outdoor gamma-ray detector into realistic source terms for later simulations.
This effort entailed the development of a full-spectrum model for all background sources in addition to the radon progeny (\Fref{sec:modeling-detector}), and the modeling was performed to a level that goes beyond other research in this area.
Instead of na\"{\i}vely inverting each measurement, we applied a simple physical model of radon progeny decay in rain droplets so that the temporal evolution of the inversion was constrained by physics (\Fref{sec:modeling-radon-progeny}).
This modeling of both the detector response and physical evolution of the radon progeny was performed on 28~separate rain events, and the results include the estimation of physically interesting quantities, such as isotope ratios and the radiometric age of the rain (\Fref{sec:results}).
Finally, the applications and significance of this work are discussed (\Fref{sec:discussion} and \Fref{sec:conclusions}), such as how the fine time resolution and the low statistical fluctuations of the method could potentially be useful for meteorological and agricultural applications in addition to the intended security application.


\section{Full-spectrum analysis of gamma-ray background}\label{sec:modeling-detector}
Across multiple fields over recent decades, there has been a general trend to utilize the full spectrum from a spectroscopic detector rather than just the counts in particular photopeaks or coarse regions of interest~\cite{minty_airborne_1992,hendriks_full-spectrum_2001,caciolli_new_2012,bandstra_modeling_2020,xu_analysis_2022}, which allows for more fully leveraging spectroscopic information.
This approach is especially useful for detectors like NaI(Tl), whose energy resolution often does not allow one to cleanly resolve certain photopeaks, but whose size and durability offer great advantages over, e.g., HPGe, in terms of efficiency, cost, and ease of use.

In this section we will describe measurements using a NaI(Tl) detector (\Fref{sec:detector-measurements}), and then describe the two-stage Monte Carlo-based simulation approach used to obtain quantitative spectra for each of the contributions to the background (\Fref{sec:detector-modeling}).
We will then explain how the full-spectrum background model was used to correct the gain of the measured spectra (\Fref{sec:gain-correction}), and finally how the activity concentrations of radon progeny were inferred from measurements (\Fref{sec:estimating-densities}).


\subsection{Static detector measurements}\label{sec:detector-measurements}
\begin{figure}
    \begin{center}
        \includegraphics[width=0.7\columnwidth]{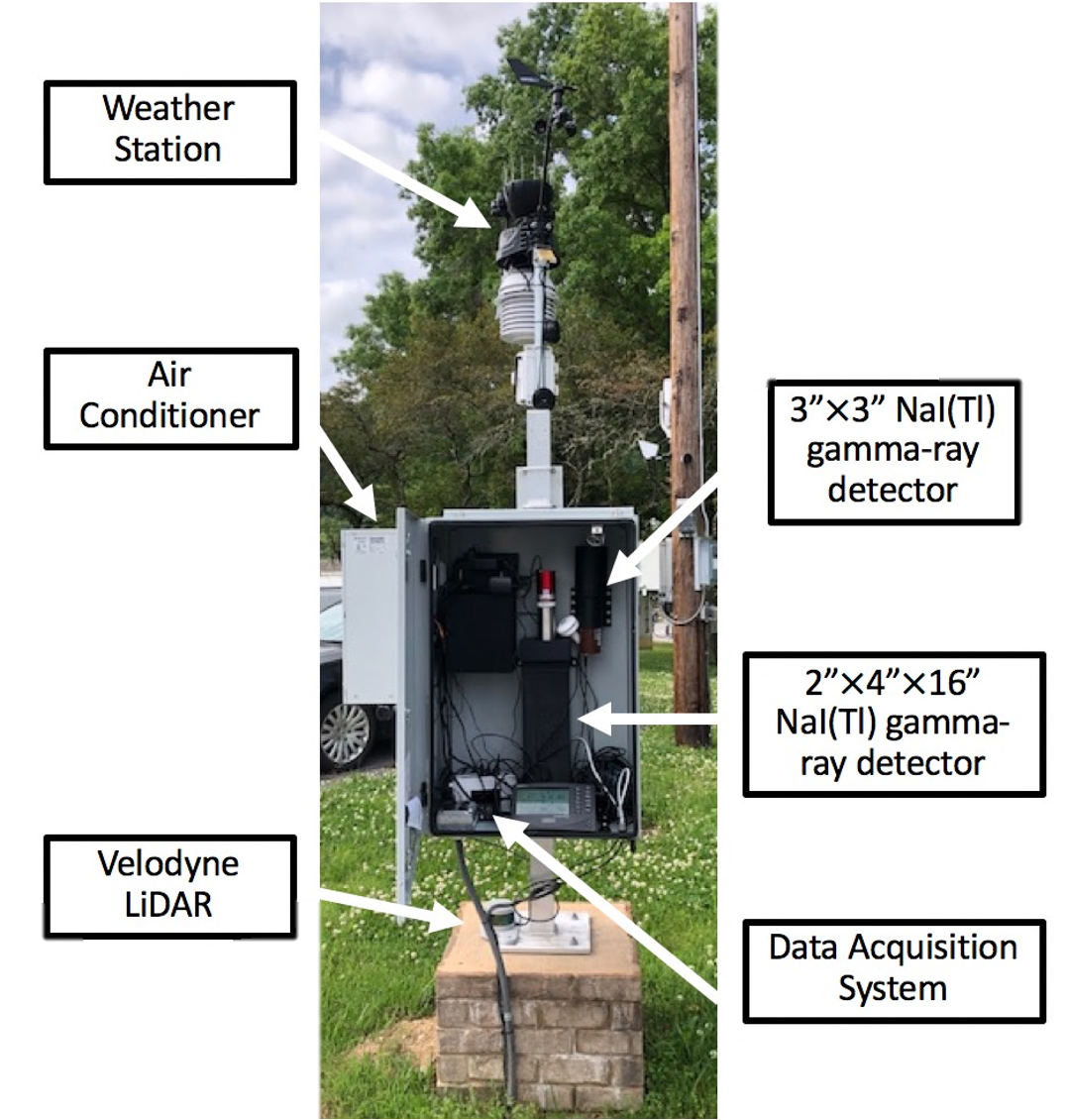}
    \end{center}
    \caption{MUSE01 node, with major components labeled. Figure from ref.~\cite{archer_radiation_2019_minos3}.\label{fig:muse_node}}
\end{figure}

Experimental data were taken from a multi-year dataset of a stationary detector system named MUSE01 located at the entrance to the High Flux Isotope Reactor / Radiochemical Engineering Development Center (HFIR/REDC) at Oak Ridge National Laboratory (ORNL)~\cite{nicholson_characterization_2018_minos1,nicholson_data_2019_minos2,archer_radiation_2019_minos3}.
The system consists of a variety of sensors including: one 2''\(\times\)4''\(\times\)16'' NaI(Tl) detector with a Mirion Osprey multichannel analyzer (MCA), one 3''\(\times\)3'' NaI(Tl) detector with an ORTEC DigiBase MCA, a Davis Vantage Pro weather station, Velodyne VLP-16 Puck LiDAR, a data acquisition computer, and HVAC (heating, ventilation, and air conditioning) for temperature stability.
The data for this work were taken from the large NaI(Tl) and the weather station.
The NaI(Tl) data were recorded in listmode, binned into 100\,ms gamma-ray energy spectra with 1000 linear bins from 0 to 3\,MeV, and uploaded to a remote server.
The weather station data were collected in 1-second intervals and contain a plethora of environmental data including rainfall rate, total rainfall accumulation, humidity, pressure, temperature, soil moisture and temperature, and solar radiation.
The data acquisition system recorded both data streams in a time aligned SQLite~3 database with a clock that is synchronized with a network time protocol (NTP) server that is, in turn, synchronized with GPS\@.
Cameras positioned around the detection system along with the production facility logs are used as ground truth as the system data are analyzed.
The system is shown in \Fref{fig:muse_node}.

Rain events for this analysis were selected by anomaly detection algorithms based on the increase in gamma ray count rate from radon washout/fallout including \isot{Bi}{214} (609, 1764 and 2204\,keV) and \isot{Pb}{214} (295 and 352\,keV).
The rain events were then confirmed with video data.
These rain events have additionally been used to evaluate radiation anomaly detection algorithms, where they have proven to be a common source of false alarms for traditional detection algorithms~\cite{ghawaly_characterization_2022}.

\subsection{Modeling the full-spectrum detector background}\label{sec:detector-modeling}
The measured spectra due to the various sources of background were simulated in the following two-stage process, which is similar to that used in ref.~\cite{sandness_accurate_2009}.
This modeling is needed in order to relate environmental quantities like activity concentrations to detector-measured quantities, also known as finding the ``conversion factors''~\cite{minato_monte_1980,nishikawa_analysis_2000}.
First a large-volume simulation of a cylindrical geometry was performed, which was used to tally particles on a hemisphere around the detector, and the second simulation performs the close-in simulation inside the hemisphere around the detector.
The first-stage geometry consisted of a cylinder of radius 300\,m and atmosphere height 300\,m and soil depth 1.25\,m.
As suggested by ref.~\cite{peplow_monte_2021}, extending the radius and atmosphere height to at least 500\,m and soil to 2\,m deep is needed to more accurately model the low-energy portion of the detector spectrum around 100\,keV\@.
Our approach differs from both in expanding the scale of the first stage by an even greater distance and exploiting more symmetries to improve the efficiency of the process.


\subsubsection{Background simulations - stage 1}\label{sec:modeling-stage1}
\begin{figure*}
\begin{center}
\input{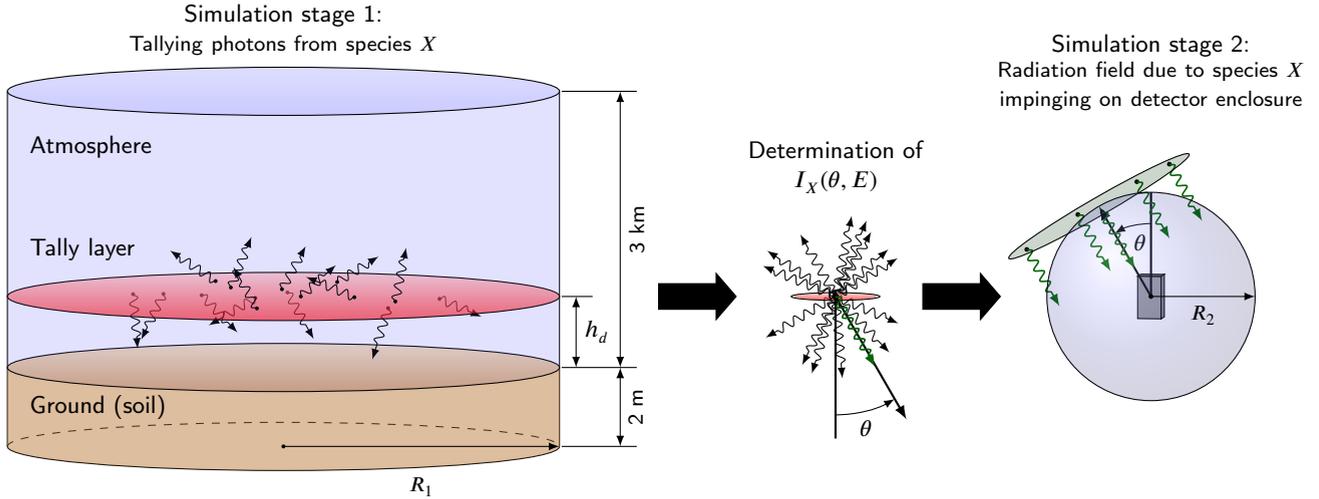}
\end{center}
\caption{Graphical depiction of the two stage Monte Carlo background simulations.\label{fig:sims-diagram}}
\end{figure*}

For the first stage of the simulations, the goal was to calculate the radiation field in all directions at a point above the ground at the same height as the eventual detector (\( h_d \)\,=\,1.3\,m).
In order to simulate a realistic background spectrum, one needs to simulate using large volumes, even extending to multiple kilometers, in order to properly include all of the downscatter from all of the background sources~\cite{peplow_monte_2021}.
The criterion we chose is that the atmosphere and soil must both be thicker than ten times the mean free path of the highest energy prominent background line at 2614\,keV, which has a mean free path of approximately 215~meters in air and 17~cm in soil~\cite{xcom}.
To efficiently simulate such a large volume, we exploited the fact that a uniform planar geometry would yield a background radiation field that is translationally symmetric, only varying as a function of altitude.
So we performed the simulation in a cylindrical geometry of radius \( R_1 \)\,=\,10\(^4\)\,km and height \( H \)\,=\,3\,km.
A cylindrical geometry was chosen so that the ``boundary region'' with less realistic downscattering was uniform (as compared to a square), and the radius \( R_1 \) was chosen to be so large (even larger than the Earth's radius) in order that the ratio of the lateral air boundary area to the tally plane area would be only approximately \( 2 \pi R_1 H / \pi R_1^2 = 2 H / R_1 \)\,<\,0.1\% of the total volume and thus could be ignored.
The cylinder contained a 2-m thick soil cylinder to represent the ground, and above it was a 3000-m atmosphere of density \(1.205 \cdot 10^{-3}\)\,g/cm\(^3\).
Like the atmosphere height, the ground thickness was chosen to be at least several mean-free paths of the highest energy emission.
A plane that tallies particles crossing it was placed at height \( h_d \).
All materials used in this and other simulations were taken from ref.~\cite{detwiler_compendium_2021}, specifically ``Air (Dry, Near Sea Level)'' and ``Earth, U.S. Average''.
The simulation geometry is depicted on the left side of \Fref{fig:sims-diagram}.

We note that other methods could be used to obtain the same tally results more efficiently, such as by using reflective boundary conditions.
This method was chosen for its simplicity in setting up and running on any Monte Carlo tool of choice.


The various background sources were divided into three different groups depending on their physical origin:

\paragraph{(1) Terrestrial and long-lived fallout sources}
These sources consist of the primordial \isot{U}{238} series, \isot{U}{235} series, \isot{Th}{232} series, and \isot{K}{40}.
The decay chains were assumed to be in equilibrium and all species were assumed to be distributed evenly throughout the ground.
Also included in this group was \isot{Cs}{137} from weapons testing fallout~\cite{ritchie_application_1990}.
It is generally distributed closer to the surface than the primordial isotopes but has had decades to diffuse into the soil.
For simplicity it was assumed to be evenly distributed as well, as has been done elsewhere~\cite{swinney_methodology_2018}.

\paragraph{(2) Atmospheric and cosmic sources}
These sources are those distributed throughout the lower atmosphere.
This group contains exactly one member, positron annihilation, which results from positrons produced by cosmic rays in the atmosphere.
Each annihilation produces two 511\,keV photons, and this source was simulated as originating evenly from every part of the atmosphere volume.
Suspended \isot{Rn}{222} progeny may also be present in the lower atmosphere and could be included here as well.
However, their presence should be largely compensated for by the initial full-spectrum fit of the background before each rain event, since their signatures resemble the \isot{U}{238} decay series.
Because of this, we have neglected them, although in principle a large change in their levels during a rain event could cause systematic errors.

\paragraph{(3) Short-lived fallout sources}
The third group are naturally occurring short-lived fallout species from the atmosphere that exists only on the surface and are primarily deposited via precipitation.
This group contains the isotopes of interest, \isot{Pb}{214} and \isot{Bi}{214}, as well as \isot{Be}{7}, a cosmogenic isotope with a single prominent photopeak at 477\,keV and a half-life of 53~days~\cite{nndc_nudat,wallbrink_use_1993,blake_fallout_1999,mabit_comparative_2008}.
These isotopes were simulated as coming from a uniform volume consisting of the top 9\,mm of the ground volume.
The determination of this depth will be described later.

\paragraph{}  
Simulations for each of these species was performed using MEGAlib's Cosima tool~\cite{zoglauer_megalib_2006,zoglauer_cosima_2009}, which is a wrapper for the Geant4 Monte Carlo code~\cite{agostinelli_geant4_2003,allison_geant4_2006,allison_geant4_2016}.
Each species was simulated with a nominal activity or emission rate in the relevant units.
For example, terrestrial \isot{K}{40} was simulated assuming an activity concentration of \( a_{\mathrm{K40}} \)\,=\,10\(^{-9}\)\,Bq/m\(^3\), fallout \isot{Pb}{214} was simulated assuming \( a_{\mathrm{Pb214}} \)\,=\,10\(^{-9}\)\,Bq/m\(^2\), and positron annihilation photons were simulated assuming an annihilation rate density of \( a_{\mathrm{ann.}} \)\,=\,10\(^{-9}\)\,annihilations/s/m\(^{3}\).
These concentration units will be referred to as Bq/m\(^{\{2,3\}}\) if the units can be determined from context.
The simulation times used were chosen to be long enough to generate \( 10^8 \) photons, which for all cases supplied at least \( 10^6 \) photons tallied by the tally plane.
The entire simulation for each species took approximately 6--10~CPU-hours on a 2\,GHz CPU, and the simulations can easily be run in parallel.
\Fref{tab:photons-per-decay} lists all of the species that were simulated, along with the units of activity concentration and the number of photons emitted per decay or annihilation event (i.e., the sum of all photon branching ratios).

\begin{table*}
    \center
    \caption{Summary of information about the various species that were simulated.\label{tab:photons-per-decay}}
    \begin{tabular}{ c c c c }
        species
        & group
        & photons per event
        & concentration units
    \\
    \midrule
        \isot{K}{40}
        & Terrestrial/long-lived fallout
        & 0.116
        & Bq/m\(^3\)
    \\
        \isot{U}{238}
        & Terrestrial/long-lived fallout
        & 3.04
        & Bq/m\(^3\)
    \\
        \isot{U}{235}
        & Terrestrial/long-lived fallout
        & 4.89
        & Bq/m\(^3\)
    \\
        \isot{Th}{232}
        & Terrestrial/long-lived fallout
        & 3.79
        & Bq/m\(^3\)
    \\
        \isot{Cs}{137}
        & Terrestrial/long-lived fallout
        & 0.929
        & Bq/m\(^3\)
    \\
        positron annihilation (511\,keV)
        & Atmospheric/cosmic
        & 2.00
        & annihilations/s/m\(^3\)
    \\
        \isot{Be}{7}
        & Short-lived fallout
        & 0.104
        & Bq/m\(^2\)
    \\
        \isot{Pb}{214}
        & Short-lived fallout
        & 0.954
        & Bq/m\(^2\)
    \\
        \isot{Bi}{214}
        & Short-lived fallout
        & 1.34
        & Bq/m\(^2\)
    \\
    \end{tabular}
\end{table*}

After the simulations were performed, the results were analyzed in the following fashion.
All of the photons that were recorded by the tally plane were binned according to their energy \( E \) and zenith angle \( \theta \), which was defined as the angle between the photon's direction of travel and the negative vertical axis, i.e., the angle that the photon appears to come from, so that a photon traveling vertically upward has a zenith angle of \( 180^{\circ} \) and a photon traveling vertically downward has a zenith angle of \( 0^{\circ} \).
All of these tallied events were then used to calculate the mean intensity of photons at the height of the tally plane.

Defining the radiation intensity \( I_{X} \) for species \( X \) in units of photons per second per keV per m\(^2\) per steradian per Bq/m\(^{\{2,3\}}\), then the relationship between the infinitesimal observed number of photons \( dN_1 \) due to a nominal activity concentration \( a_{X} \) crossing an infinitesimal area \( dA_1 \) of the tally plane during observation time \( dT_1 \), in an energy bin of width \( dE \) around the photon energy \( E \), and into an infinitesimal solid angle \( d\Omega \) around zenith angle \( \theta \) is
\begin{align}
    dN_1
        &=
            I_{X}(E, \theta)
            \cdot
            dT_1
            \cdot
            dE
            \cdot
            dA_1
            \cdot
            \cos(\theta)
            \cdot
            d\Omega
            \cdot
            a_{X}
        ,
\end{align}
where the cosine term arises from the fact that the photons are tallied crossing a plane and thus the projected area \( dA_1 \cdot \cos(\theta) \) is the relevant area.

Integrating the infinitesimals across the entire tally plane and simulation time and into non-infinitesimal energy and zenith angle bins, we get the mean intensity for species \( X \) in those bins
\begin{align}
    I_{X}(E, \theta)
        &\approx
            \frac{
                N_1
             }{
                T_1
                \cdot
                \Delta E
                \cdot
                \pi R_1^2
                \cdot
                \pi \left[
                    \cos^2\left(
                        \theta - \frac{\Delta\theta}{2}
                    \right)
                    -
                    \cos^2\left(
                        \theta + \frac{\Delta\theta}{2}
                    \right)
                \right]
                \cdot
                a_{X}
             }
        .
\end{align}

\begin{figure*}
    \begin{center}
        \begin{tikzpicture}
            \draw (0, 0) node[inner sep=0] {\includegraphics[width=0.7\columnwidth]{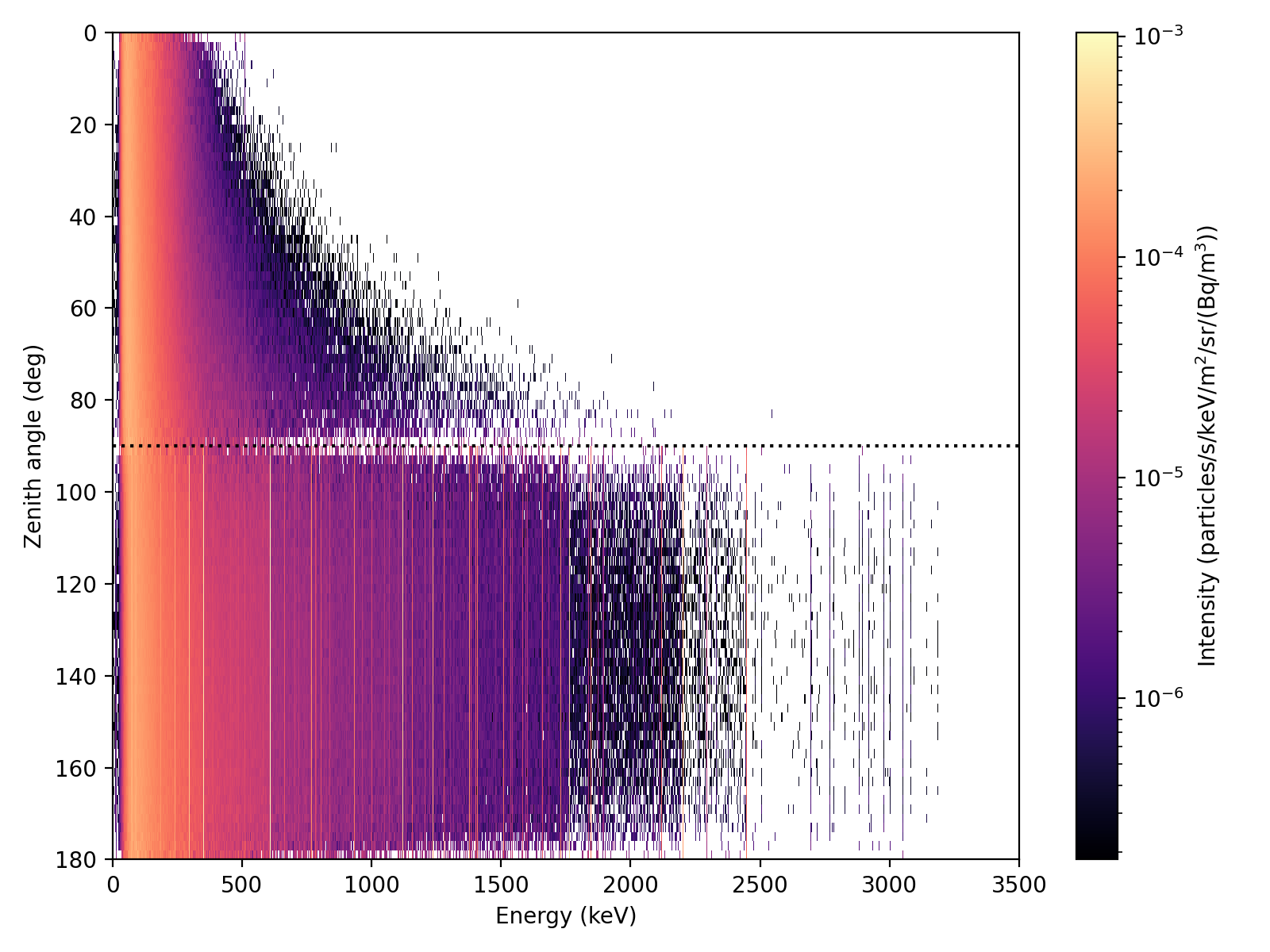}};
            \draw (2.3, 3.3) node {\Large \textbf{\isot{U}{\mathbf{238}}}};
        \end{tikzpicture}\\
        \begin{tikzpicture}
            \draw (0, 0) node[inner sep=0] {\includegraphics[width=0.7\columnwidth]{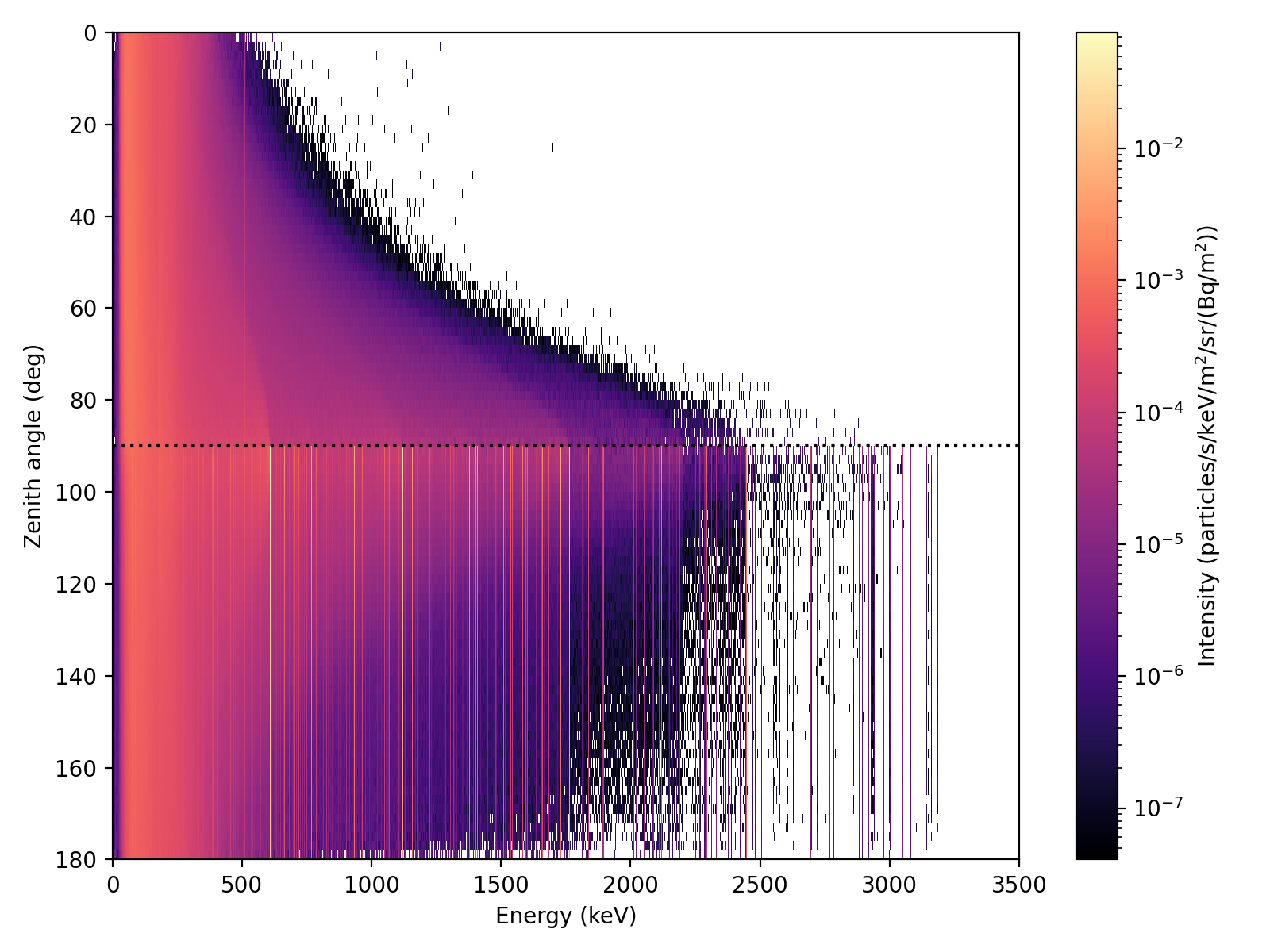}};
            \draw (2.3, 3.3) node {\Large \textbf{\isot{Bi}{\mathbf{214}}}};
        \end{tikzpicture}
    \end{center}
    \caption{Radiation field intensity for the \isot{U}{238} series terrestrial source (top) and \isot{Bi}{214} surface source (bottom) calculated during the first stage of the simulation.\label{fig:sim-stage1-intensity}}
\end{figure*}

Photon events were binned in bins of width 3\,keV from 0 to 3501\,keV and zenith angle bins of width 2\(^{\circ}\) from 0 to 180\(^{\circ}\).
Examples of such intensity determinations for the \isot{U}{238} series and \isot{Bi}{214} are shown in~\Fref{fig:sim-stage1-intensity}.
Those two sources have many of the same lines (\isot{Bi}{214} is in the \isot{U}{238} series and contributes the majority of the series' strongest lines), but their intensities show some major differences.
For the \isot{U}{238} series, the intensity of upward-traveling photons (the lower half of the plot) is approximately constant as a function of zenith angle.
This effect is the result of both the uniformity of terrestrial sources in the soil and the shielding of the soil.
Specifically, for a point just above the surface, the photons that reach it traveling upward at a certain angle on average come from the same depth, which is the mean free path of the photon's energy in soil.
Assuming the soil is much thicker than the mean free path, then the intensity should be a constant for all upward-going directions at a given energy.
By contrast, the \isot{Bi}{214} intensity is more strongly peaked at a zenith angle of 90~degrees for most photon energies.
This observation agrees with the fact that the source is evenly distributed over the surface of the soil, so emission from that layer that directly reaches the detector would have a zenith angle of approximately 90~degrees.
A commonality between both sources is that photopeaks are only seen in emission from below, and there are no sharp features in the emission coming from above.
This agrees with the fact that a photon that arrives from above would have necessarily had to have scattered in the air first.

\subsubsection{Background simulations - stage 2}\label{sec:modeling-stage2}
The second stage of the simulations consisted of taking the radiation field intensity for each background source type and using it to recreate the same field inside a spherical volume of radius \( R_2 \)\,=\,0.8\,m centered around the detector.
The radius was chosen to surround the entire system enclosure, and the system enclosure and detector were modeled based on measurements of the system and validated using calibration source measurements.

To perform these simulations, each tally bin centered at \( (E, \theta) \) with widths \( (\Delta E, \Delta \theta) \) were converted into a ``far-field'' source of photons incident on the sphere with zenith angles \( \theta \), uniformly distributed azimuth angles, energy \( E \), and a mean number of generated photons equal to
\begin{align}
    N_2(E, \theta)
        &=
            I_{X}(E, \theta)
            \cdot
            T_2
            \cdot
            \Delta E
            \cdot
            \pi R_2^2
            \cdot
            \pi \left[
                \cos^2\left(
                    \theta - \frac{\Delta\theta}{2}
                \right)
                -
                \cos^2\left(
                    \theta + \frac{\Delta\theta}{2}
                \right)
            \right]
            \cdot
            a_{X}
        .
\end{align}

Once again, simulations were run for each of the sources.
The same nominal activity concentrations were assumed, and the simulation times \( T_2 \) were chosen to be long enough to generate \( 10^9 \) photons, which led to on average \( 10^7 \) detected events for each source.
The simulation for each source took approximately 8--10~CPU-hours on a single-core 2\,GHz CPU though it was easily broken into smaller pieces and parallelized.

The result of the second stage of simulations is a spectrum \( S(E) \) in units of events per second per keV per Bq/m\(^{\{2,3\}}\), obtained by tallying the events into energy bins \( \Delta E \) around each \( E \), and dividing by \( a_{X} \cdot \Delta E \cdot T_2 \).

This stage of the Monte Carlo simulations is depicted on the right side of \Fref{fig:sims-diagram}.


\subsubsection{Validation of the simulations}\label{sec:modeling-validation}
To validate the model, a 10-minute long spectrum was taken from one of the rain events when there had been no rain for over a day.
This spectrum was used to fit a model that consisted of a linear combination of all of these source spectra.
Each simulated spectrum is multiplied by its coefficient, the observation time, and energy bin widths to obtain the mean counts from that source.
Therefore the coefficients are the activity concentrations \( a_{X} \) in Bq/m\(^{\{2,3\}}\) for each of the sources.
The \isot{U}{235} decay chain coefficient was tied to the \isot{U}{238} coefficient assuming their mean natural abundances, i.e., a ratio of \isot{U}{235} to \isot{U}{238} activity of 0.04605.

One additional source that was not simulated herein was also included.
This was the cosmic-ray continuum, for which we used a heuristic model.
This continuum has previously been modeled as a power law, but departs from a pure power law at low energies~\cite{sandness_accurate_2009,bandstra_correlations_2021}.
Here we used an exponential cutoff power law, typically expressed as \( E^{-\alpha} \exp(-E/\beta) \), which suppresses the high energy portion of the spectrum~\cite{arnaud_xspec_1996,clauset_power-law_2009}.
To instead suppress the low energy portion of the spectrum, we modified the function so that the inverse of the energy was used in the exponent.
Recasting the formula so that the additional parameter is the energy at which the function reaches its maximum, we obtain
\begin{align}
    S(E)
        &=
            A_{\mathrm{max}}
            \;
            {\left(
                \frac{E}{E_{\mathrm{max}}}
            \right)}^{-\alpha}
            \;
            \exp\left[
                \alpha
                \left(
                    1
                    -
                    \frac{E_{\mathrm{max}}}{E}
                \right)
            \right]
        ,
\end{align}
where the power-law index \( \alpha \) must be greater than zero.
All three parameters were allowed to fit to the data.

\begin{figure*}
    \begin{center}
        \includegraphics[width=0.60\columnwidth]{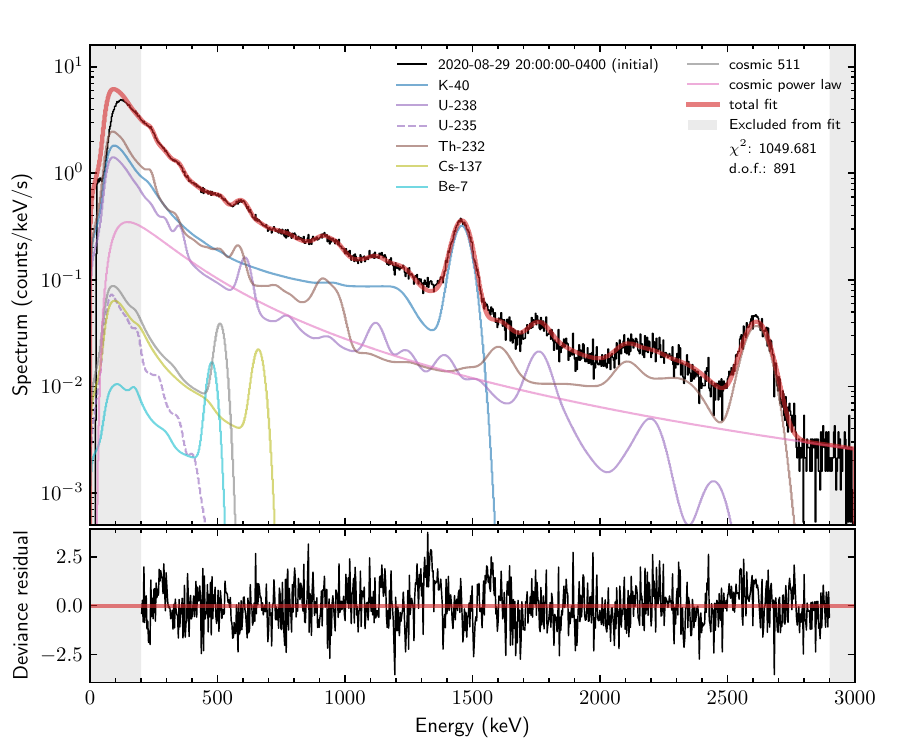}
        \includegraphics[width=0.60\columnwidth]{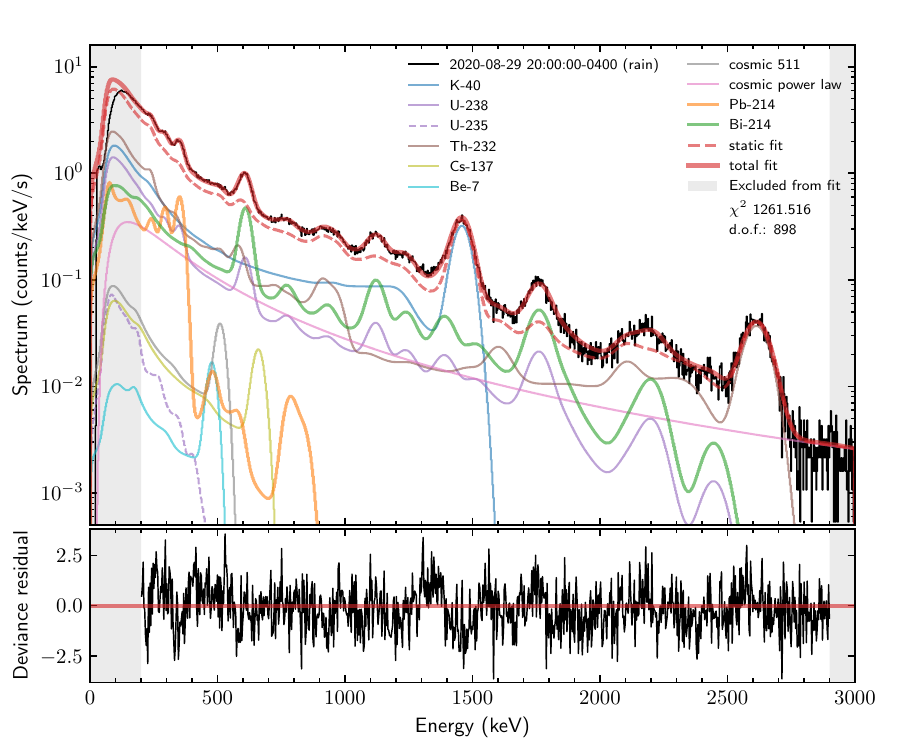}
    \end{center}
    \caption{Validation of the simulation-derived static background model with a fit to data before rain occurred (top) and during rain with radon progeny components added while the static background was held fixed (bottom). Both fits show the overall fit statistic \( \chi^2 \) and the degrees of freedom (d.o.f.)\@.\label{fig:sim-stage2-spectra}}
\end{figure*}

The total model including this cosmic continuum was fit by minimizing the Poisson negative log likelihood, and the fit is shown in the top plot of~\Fref{fig:sim-stage2-spectra}.
The spectral regions below 200\,keV and above 2900\,keV were excluded from the fit.
The low-energy region was excluded because it is most sensitive to differences between the Monte Carlo geometry and reality arising due to the strong effects of attenuation and scattering in that portion of the gamma-ray spectrum.
The high-energy region was excluded since corrections for gain shifts (\Fref{sec:gain-correction}) sometimes resulted in the upper threshold at 3\,MeV being effectively shifted downward into that portion of the spectrum.

To show that the fit is close across the entire spectrum, the deviance residuals are plotted underneath the spectrum in the top plot of~\Fref{fig:sim-stage2-spectra}.
Deviance residuals are the generalization of the \( \chi \) statistic, i.e., the difference between the measurement and prediction normalized by the standard deviation.
For Poisson statistics this residual is defined as
\begin{align}
    \chi(n, \hat{n})
        &\equiv
            \sqrt{
                2
                \left(
                    -(n - \hat{n})
                    +
                    n \log\left(\frac{n}{\hat{n}}\right)
                \right)
            }
        ,
\end{align}
where \( n \) are the measured counts and \( \hat{n} \) are the counts predicted by a model~\cite{mccullagh_generalized_1989}.

For the fits shown, these residuals do not exceed the equivalent of 3 Gaussian ``sigmas''.
However, the overall fit statistic (\( \chi^2 \), i.e., the sum of the squared deviance residuals, which is also twice the negative log likelihood) is 1050 with 891~degrees of freedom, which means it is not a statistically acceptable fit (\(p\)\,=\,1.9\(\times\)10\(^{-4}\)).
This apparent tension can be resolved by noting that the residuals are small, but there are some systematic patterns in the residuals that are not fully modeled.
Future work would be needed to further improve the accuracy of the model.
In fact, earlier versions of this model excluded \isot{Cs}{137} and \isot{Be}{7}, but small deviations in the 662\,keV and 477\,keV regions pointed to their possible presence.
The fit improved significantly when both were included, with the fit statistic decreasing from 1082 to 1055 with the addition of \isot{Cs}{137}, and from 1055 to 1050 with the addition of \isot{Be}{7}.

The concentrations obtained for the various species are shown in~\Fref{tab:fit-results}.
The results for the terrestrial sources \isot{K}{40}, \isot{U}{238}, and \isot{Th}{232} are all consistent with results that have been reported elsewhere~\cite{trevisi_natural_2012,swinney_methodology_2018,salathe_determining_2021}.
The results for \isot{Cs}{137} are also on the same order as other published results~\cite{wallbrink_use_1993,swinney_methodology_2018}.

The standard deviation of the \isot{Be}{7} results is less meaningful than the others since it is a rain-deposited cosmogenic nuclide with a 53-day half-life, so it is not expected to be constant.
Notably, its range is consistent with other published activity concentrations, which tend to be in the hundreds of Bq/m\(^2\) and strongly depends on the history of rainfall at the site~\cite{blake_fallout_1999}.

The cosmic power-law index of approximately 2.2 is larger than the values of 1.3 or lower reported elsewhere~\cite{sandness_accurate_2009,bandstra_modeling_2020}, even though those other measurements involved detectors that are similar in size to this one.
The difference could indicate a deficiency with the cutoff power-law model. Without any data above 3\,MeV, it is difficult to resolve the question.

The validation of the model was limited to its being consistent with the measured spectra and other published results.
Unfortunately, no ground truth measurements were available for the site.
Also, the site was near a road, trees, and uneven terrain, so measuring ground truth would be an involved undertaking.
Instead, we conclude that the model reflects plausible values for the environment, but not necessarily the actual values.

\begin{table*}
    \center
    \caption{The results of fits to static background data.\label{tab:fit-results}}
    \begin{tabular}{ c c c c c c }
    source/parameter  &  min.      &  mean      &  max.      & std.\ dev. & units \\
    \midrule
    \isot{K}{40}      &  \(  466            \) &  \(  540            \) &  \(  615            \) &  \(   37.7            \) & Bq/kg        \\
    \isot{U}{238}     &  \(   17.7            \) &  \(   21.7            \) &  \(   29.7            \) &  \(    2.42            \) & Bq/kg        \\
    \isot{U}{235}     &  \(    0.82            \) &  \(    1.00            \) &  \(    1.37            \) &  \(    0.11            \) & Bq/kg        \\
    \isot{Th}{232}    &  \(   25.5            \) &  \(   27.6            \) &  \(   30.8            \) &  \(    1.63            \) & Bq/kg        \\
    \isot{Cs}{137}    &  \(    0.38            \) &  \(    1.64            \) &  \(    3.10            \) &  \(    0.59            \) & Bq/kg        \\
    \isot{Be}{7}      &  \(  186            \) &  \(  743            \) &  \( 1182            \) &  \(  247            \) & Bq/m\(^2\)   \\
    positron annihilation &  \(    0.84            \) &  \(    1.52            \) &  \(    2.63            \) &  \(    0.43            \) & annihilations/s/m\(^3\)   \\
    cosmic \(A_{\mathrm{max}}\) &  \(  0.94 \cdot 10^{-1} \) &  \(  2.81 \cdot 10^{-1} \) &  \(  9.60 \cdot 10^{-1} \) &  \(  1.67 \cdot 10^{-1} \) & counts/keV/s \\
    cosmic \(E_{\mathrm{max}}\) &  \(    87               \) &  \(   159               \) &  \(   274               \) &  \(    41               \) & keV          \\
    cosmic \(\alpha\) &  \(    1.96            \) &  \(    2.22            \) &  \(    2.49            \) &  \(    0.13            \) & none
    \end{tabular}
\end{table*}


\subsubsection{Estimation of surface deposit thickness}\label{sec:thickness-estimation}
Earlier it was mentioned that the \isot{Pb}{214}, \isot{Bi}{214}, and \isot{Be}{7} sources emerged uniformly from the top 9\,mm of ground in the first stage of simulations.
The thickness of this emitting layer was determined in the following fashion.
Various thicknesses were used in stage~1, ranging from 1\,mm to 20\,mm, and the resulting spectra from both stages of the simulation were stored and fit against multiple spectra measured during intense rainfall.
The best overall figure of merit was obtained for a thickness of 9\,mm, which is shown in the bottom plot of~\Fref{fig:sim-stage2-spectra}.
This thickness changes the scattering and attenuation of the emission in the soil, which affects the measured peak-to-total ratios of the gamma-ray lines as well as the overall strength of the lines --- specifically, the thicker the layer, the lower the peak-to-total ratio and the lower the efficiency.

We expect this layer is needed to mimic the effects of vegetation and other surface roughness that these isotopes experience in the environment before reaching the detector.
Similar surface effects were noted by ref.~\cite{minato_monte_1980} as an unknown systematic uncertainty when finding conversion factors for radon progeny from Monte Carlo simulations.
One of the benefits of full-spectrum modeling is that the combination of photopeaks and continuum can serve as a consistency check that simulations of photopeak emission alone do not afford.


\subsection{Calibration and gain correction}\label{sec:gain-correction}

The spectra are post-processed first using a coarse automated gain stabilization procedure (Method~A) and subsequently by a finer method (Method~B).
The goal of both methods is to provide as accurate a channel-to-energy calibration as possible over time, which is needed because temperature affects the response of scintillator detectors and photomultiplier tubes.

Method~A periodically recalibrates the energy scale by identifying known environmental gamma-ray peaks in the summed spectrum.
Specifically, every 600 seconds (10 minutes) of data acquisition, a calibration spectrum is constructed by summing the preceding 10 minutes of measurements.
This spectrum is smoothed using a Savitzky-Golay filter, and prominent photopeaks are detected using a prominence-based peak finding algorithm.

The 1460.8\,keV line from \isot{K}{40} is used as a calibration anchor.
This peak is fitted using a Gaussian model combined with a linear background term, and peak centroids are extracted with sub-channel precision via constrained non-linear least squares optimization.
Using the fitted centroid, the detector's energy calibration is reconstructed by applying a detector-specific nonlinearity correction (via a precomputed spline).
This yields a channel-to-energy mapping that is updated every 600 seconds.

Each raw spectrum is then rebinned to a fixed global energy axis by resampling.
Individual counts in each original histogram bin are uniformly redistributed across that bin's energy width and re-aggregated into the target energy bins.
This preserves total counts and avoids interpolation bias.

If no valid calibration peak is detected, the last known good calibration is reused until the next successful recalibration.
While this approach has proven to be robust under most operating conditions, precipitation events can lead to errors in the correction due to distortion of the target photopeak.
This effect happens especially during heavy precipitation events, which is likely due to strong \isot{Bi}{214} lines appearing in the neighborhood of the 1460\,keV peak, and this can have significant effects on the calibration quality.

Method~B improves upon the results of Method~A by fitting features across the entire spectrum, specifically by exploiting the full spectrum model developed in previous sections.

To begin, the spectral background model without \isot{Pb}{214} and \isot{Bi}{214} was first fit to 10~minutes of data for each event before any rain had occurred.
This was performed as a quality check for the rain event to make sure it had no obvious nuisance sources.
The overall shape of that fitted spectral model was frozen, i.e., the different isotopes were no longer allowed to change relative to each other, and this shape was used for subsequent fits as the ``static'' background.

For every 60~seconds of the rain event, the data were stochastically rebinned assuming slight relative gain shifts around the calibration produced by Method~A\@.
We used 100 different relative gains between 98\% and 102\% of the current gain.
Each rebinned spectrum was fit with a linear model consisting of three spectral components: the static background, \isot{Pb}{214}, and \isot{Bi}{214}.
The figure of merit (negative log likelihood) was stored for each fit, and a quadratic polynomial was fit to the results of all of the trials.
The relative gain that minimized the polynomial was chosen as the best gain correction for those 60~seconds of data and was stored.
If the minimum was at the edge of the range of relative gains, then the same procedure was done starting with a wider range of gains, and this was repeated until a minimum inside the interval is found.
This method is similar to a more complex full-spectrum calibration procedure discussed elsewhere~\cite{salathe_real-time_2020}.

\begin{figure*}
    \begin{center}
        \includegraphics[width=0.95\textwidth]{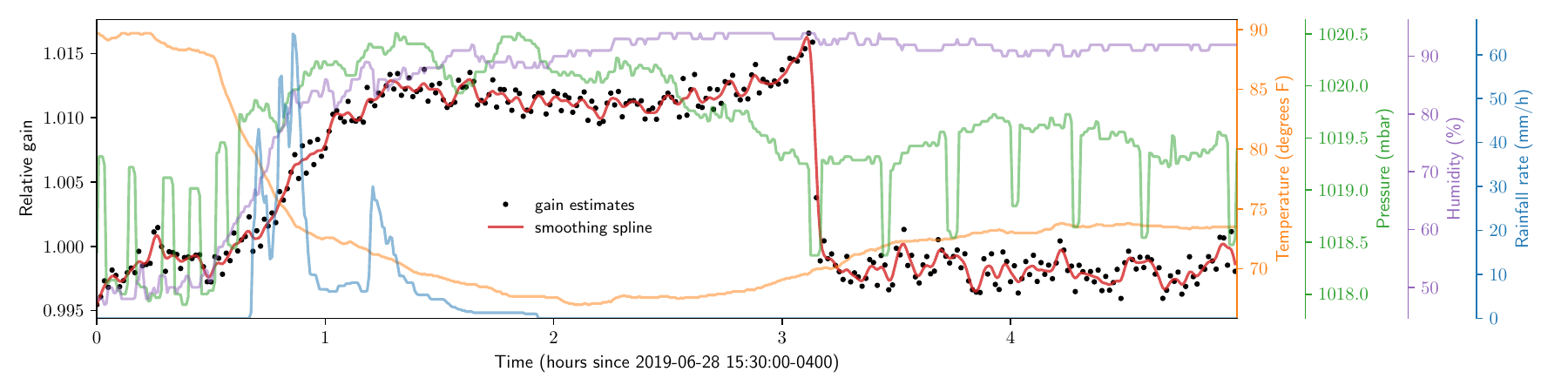}
    \end{center}
    \caption{The estimated gain adjustments (black points) and the smoothing spline used to interpolate them for one of the rain events. Also shown are various data from the weather station. The frequent drops in the pressure readings are an instrument issue.\label{fig:gain-correction}}
\end{figure*}

This procedure was performed on every 60~seconds of data through each rain event.
The resulting set of optimized gain values were fit with a smoothing spline and used in the next and final portion of the analysis.
An example of the gain correction for one rain event is shown in~\Fref{fig:gain-correction}.
The relative gain correction that Method~B produces generally drifts gradually over time, though sometimes there can be abrupt changes in the gain, as occurs at about 3.1~hours, which is likely compensating for Method~A's losing track of the \isot{K}{40} line centroid and resetting.

\subsection{Estimating radon progeny activity concentrations}\label{sec:estimating-densities}
The final step of the spectral analysis consisted of extracting the estimated \isot{Pb}{214} and \isot{Bi}{214} activity concentrations from the data.
Every 30~seconds of data were summed into a spectrum and analyzed.
The spectrum was stochastically rebinned after its gain was corrected using the interpolated value from the previous step.
The corrected spectrum was fit with the linear model consisting of the frozen static background spectrum, the \isot{Pb}{214} spectrum, and the \isot{Bi}{214} spectrum.
The static background spectrum coefficients were frozen during the fit, so the only coefficients of the linear model were the two coefficients for \isot{Pb}{214} and \isot{Bi}{214}.
Since these spectra carried their usual units, their coefficients were their activity concentrations in units of Bq/m\(^2\), which are the final physical quantities we are trying to obtain.

The linear model fit of each spectrum \( \mathbf{x} \) to obtain the coefficients \( \mathbf{a} \) can be formulated in the following way.
Calling the static background \( \mathbf{x}_{0} \) and taking the \isot{Pb}{214}, and \isot{Bi}{214} spectra and arranging them into the columns of a matrix \( \mathbf{W} \), we obtain
\begin{align}
    \mathbf{x}
        &\approx
            \mathbf{x}_{0}
            +
            \mathbf{W}
            \mathbf{a}
        ,
\end{align}
and the maximum likelihood fit of the coefficients is
\begin{align}
    \hat{\mathbf{a}}
        &\equiv
            \argmax_{\mathbf{a} \ge 0} \log L(\mathbf{a} | \mathbf{x})
        \\
        &=
            \argmin_{\mathbf{a} \ge 0}
                \sum \left(
                    \mathbf{x}_{0}
                    +
                    \mathbf{W} \mathbf{a}
                    -
                    \mathbf{x}
                    \odot
                    \log\left(
                        \mathbf{x}_{0}
                        +
                        \mathbf{W} \mathbf{a}
                    \right)
                \right)
        .
\end{align}
Because the fit uses maximum likelihood, we can calculate the Fisher information to estimate the covariance of the coefficients:
\begin{align}
    \mathrm{cov}[\hat{\mathbf{a}}]
        &\approx
            {\left[
                \mathbf{W}^{\top}
                \mathrm{diag}\left(
                    \frac{
                        \mathbf{1}
                    }{
                        \mathbf{x}_{0}
                        +
                        \mathbf{W} \hat{\mathbf{a}}
                    }
                \right)
                \mathbf{W}
            \right]}^{-1}
        \equiv
            \boldsymbol{\Sigma}
        ,
\end{align}
which will be useful in later analysis.

In summary, through the methods in this section, we obtained the estimates of the activity concentrations of \isot{Pb}{214} and \isot{Bi}{214}, as well as a covariance matrix at a stride of 30~seconds over the course of all of the rain events.


\section{Modeling radon progeny evolution}\label{sec:modeling-radon-progeny}
Various authors use different models for the evolution of radon progeny starting from the decay of \isot{Rn}{222} in the cloud, to the attachment of the progeny to droplets, to the formation of droplets into raindrops, and the falling of raindrops to the ground.
All models agree that as soon as \isot{Rn}{222} decays to \isot{Po}{218}, the \isot{Po}{218} ion attaches to a droplet on a fast enough timescale to be considered instantaneous~\cite{priebsch_uber_1932}.
Therefore, there are virtually no free radon progeny ions in the air that have not already been attached to cloud droplets.

Models differ in how they handle the droplet phase.
Some assume the progeny attached to droplets are able to reach secular equilibrium, i.e., progeny activity ratios of 1:1:1~\cite{greenfield_determination_2008,mercier_increased_2009}.
Others assume the droplets are in equilibrium with radon for some fixed time, or else a random process with a mean lifetime, until they are removed from equilibrium and condensed into raindrops~\cite{bhandari_study_1963,minato_simple_2007,moriizumi_214bi214pb_2015}.
Still others have modeled the droplets as being instantaneously removed from equilibrium after \isot{Po}{218} attaches, with activities starting in the ratio 1:0:0~\cite{horng_rainout_2003}.

We tried multiple models of the droplet phase, from assuming secular equilibrium, to allowing for a mean removal time (with the removal time as an additional parameter of the model), to assuming instantaneous removal.
We found that secular equilibrium was generally not a good assumption and led to poor fits of the data, while the other two led to adequate fits.
For our data, the model with a finite removal time led to solutions that were largely insensitive to that time, so we concluded that assuming instantaneous removal would be both adequate and parsimonious, though it would tend to overestimate the actual age of the raindrops~\cite{moriizumi_214bi214pb_2015}.
This part of the model is described in~\Fref{sec:evolution-in-cloud}.

The droplets then condense into raindrops, which grow until they are large enough to fall to the ground.
Some models treat this process using multiple steps~\cite{damon_natural_1954,takeuchi_rainout-washout_1982}, while others consider it as one phase~\cite{horng_rainout_2003,minato_simple_2007}.
We opted again for the more parsimonious option and assumed a single ``apparent age'' for the raindrops (\Fref{sec:evolution-cloud-to-ground}).

As opposed to systems that collect periodic samples and measure them in shielded enclosures, our system is of the \textit{in situ} variety, since the detector is constantly exposed to the accumulation of all rainwater on the ground.
Thus, unlike some models, we must also consider a final phase where the progeny activity concentrations evolve while they are on the ground (\Fref{sec:evolution-on-ground}).


\subsection{Notation}\label{sec:notation}
Throughout this section we will use the shorthand of A, B, and C to stand for \isot{Po}{218}, \isot{Pb}{214}, and \isot{Bi}{214}, respectively, following their original names Radium~A, Radium~B, and Radium~C (e.g., ref.~\cite{rutherford_radioactivity_1905}).
Their decay constants are \( \lambda_{\mathrm{A}} \)\,=\,3.7290\(\times\)10\(^{-3}\)\,s\(^{-1}\), \( \lambda_{\mathrm{B}} \)\,=\,4.2692\(\times\)10\(^{-4}\)\,s\(^{-1}\), and \( \lambda_{\mathrm{C}} \)\,=\,5.8053\(\times\)10\(^{-4}\)\,s\(^{-1}\)~\cite{nndc_nudat}, which we will collectively refer to as the vector \( \pmb{\lambda} = {\left[\begin{array}{ccc} \lambda_{\mathrm{A}} & \lambda_{\mathrm{B}} & \lambda_{\mathrm{C}} \end{array}\right]}^{\top} \).

At different points it will be useful to refer to either number densities (\( n \) or \( \mathbf{n} \)) or activity concentrations (\( \alpha \), \( a \), \( \boldsymbol{\alpha} \), or \( \mathbf{a} \)).
To convert between the two, we use the relationship \( \alpha_X = \lambda_X n_X \) for each species \( X \).
Often the number densities and activity concentrations will be grouped into length-3 vectors (e.g., \( \mathbf{n} \) and \( \mathbf{a} \)), which are related by
\begin{align}
    \mathbf{a}
        &=
            \mathrm{diag}\left(
                \pmb{\lambda}
            \right)
            \cdot
            \mathbf{n}
        \\
    \mathbf{n}
        &=
            \mathrm{diag}{\left(
                \pmb{\lambda}
            \right)}^{-1}
            \cdot
            \mathbf{a}
        ,
\end{align}
where \( \mathrm{diag} \) converts a column vector into a diagonal matrix.



\subsection{Evolution of progeny inside cloud}\label{sec:evolution-in-cloud}
Assuming instantaneous attachment of \isot{Po}{218} onto cloud droplets and instantaneous removal of cloud droplets into raindrops means that the number density of \isot{Po}{218} in the cloud obeys the following relationship
\begin{align}
    \dot{n}^{c}_{\mathrm{A}}
        &=
            -
            \lambda_{\mathrm{A}}
            n^{c}_{\mathrm{A}}
            +
            \alpha_{\mathrm{Rn}}
        =
            0
        ,
\end{align}
which, assuming equilibrium has been reached, means that it is proportional to the radon activity:
\begin{align}
    n^{c}_{\mathrm{A}}
        &=
            \lambda_{\mathrm{A}}^{-1}
            \alpha_{\mathrm{Rn}}
        .
\end{align}

In terms of the activity concentration,
\begin{align}
    \alpha^{c}_{\mathrm{A}}
        &=
            \alpha_{\mathrm{Rn}}
        .
\end{align}
In other words, our assumptions amount to assuming secular equilibrium between radon and \isot{Po}{218}.


\subsection{Evolution of progeny from cloud droplets to raindrops to ground}\label{sec:evolution-cloud-to-ground}
After removal, the cloud droplets therefore begin at time \( t = 0 \) with activity concentrations in cloud air equal to
\begin{align}
    \left[
        \begin{array}{c}
            \alpha^{c}_{\mathrm{A}}(0) \\
            \alpha^{c}_{\mathrm{B}}(0) \\
            \alpha^{c}_{\mathrm{C}}(0)
        \end{array}
    \right]
        &\equiv
            \boldsymbol{\alpha}^{c}(0)
        =
            \alpha_{\mathrm{Rn}}(0)
            \,
            \mathbf{e}_{100}
        ,
\end{align}
where \( \mathbf{e}_{100} = {\left[\begin{array}{ccc} 1 & 0 & 0 \end{array}\right]}^{\top} \).

The droplets coalesce to form into raindrops, and the activity concentrations and number densities then must be measured in terms of volume of rainwater rather than volume of cloud air.
To do this, we will assume some conversion factor \( \kappa \) to scale the cloud volume densities to rainwater densities
\begin{align}
    \boldsymbol{\alpha}
        &=
            \kappa
            \boldsymbol{\alpha}^{c}
        \\
    \mathbf{n}
        &=
            \kappa
            \mathbf{n}^{c}
        ,
\end{align}
and we will leave the estimation of \( \kappa \) to be done elsewhere (e.g., ref.~\cite{minato_simple_2007}).

Call time \( t = \tau \) when the raindrops arrive at the ground, i.e., \( \tau \) is their apparent radiometric age, which can be as long as 100~minutes since it includes both the time to form raindrops and the time to fall to the ground~\cite{greenfield_determination_2008,moriizumi_214bi214pb_2015}.
From time \( 0 \) to \( \tau \) the number densities in the raindrops evolve according to the usual laws of successive radioactive decay~\cite{rutherford_radioactivity_1905}
\begin{align}
    \dot{\mathbf{n}}(t)
        &=
            -
            \left[\begin{array}{ccc}
                        \lambda_{\mathrm{A}}
                    &
                        0
                    &
                        0
                \\
                        -\lambda_{\mathrm{A}}
                    &
                        \lambda_{\mathrm{B}}
                    &
                        0
                \\
                        0
                    &
                        -\lambda_{\mathrm{B}}
                    &
                        \lambda_{\mathrm{C}}
            \end{array}\right]
            \cdot
            \mathbf{n}(t)
        ,
\end{align}
which, expressed in terms of activity concentration, is
\begin{align}
    \dot{\boldsymbol{\alpha}}(t)
        &=
            -
            \left[\begin{array}{ccc}
                        \lambda_{\mathrm{A}}
                    &
                        0
                    &
                        0
                \\
                        -\lambda_{\mathrm{B}}
                    &
                        \lambda_{\mathrm{B}}
                    &
                        0
                \\
                        0
                    &
                        -\lambda_{\mathrm{C}}
                    &
                        \lambda_{\mathrm{C}}
            \end{array}\right]
            \cdot
            \boldsymbol{\alpha}(t)
        \\
        &\equiv
            -
            \boldsymbol{\Lambda}
            \cdot
            \boldsymbol{\alpha}(t)
        .
\end{align}

This ordinary differential equation has the solution
\begin{align}
    \boldsymbol{\alpha}(t)
        &=
            \exp\left(
                -
                \boldsymbol{\Lambda}
                t
            \right)
            \cdot
            \boldsymbol{\alpha}(0)
        ,
\end{align}
and it can be efficiently calculated by finding the eigenvector diagonalization of \( \boldsymbol{\Lambda} \), which is given by
\begin{align}
    \boldsymbol{\Lambda}
        &=
            \mathbf{P}
            \cdot
            \mathrm{diag}
            \left(
                \pmb{\lambda}
            \right)
            \cdot
            \mathbf{P}^{-1}
        ,
\end{align}
where
\begin{align}
    \mathbf{P}
        &=
            \left[\begin{array}{ccc}
                        1
                    &
                        0
                    &
                        0
                \\
                        -\frac{
                            \lambda_{\mathrm{B}}
                        }{
                            (\lambda_{\mathrm{A}} - \lambda_{\mathrm{B}})
                        }
                    &
                        1
                    &
                        0
                \\
                        \frac{
                            \lambda_{\mathrm{B}} \lambda_{\mathrm{C}}
                        }{
                            (\lambda_{\mathrm{A}} - \lambda_{\mathrm{B}}) (\lambda_{\mathrm{A}} - \lambda_{\mathrm{C}})
                        }
                    &
                        -\frac{
                            \lambda_{\mathrm{C}}
                        }{
                            (\lambda_{\mathrm{B}} - \lambda_{\mathrm{C}})
                        }
                    &
                        1
            \end{array}\right]
\end{align}
is the matrix whose columns are the eigenvectors (up to an arbitrary scale factor),
and its inverse is
\begin{align}
    \mathbf{P}^{-1}
        &=
            \left[\begin{array}{ccc}
                        1
                    &
                        0
                    &
                        0
                \\
                        \frac{
                            \lambda_{\mathrm{B}}
                        }{
                            (\lambda_{\mathrm{A}} - \lambda_{\mathrm{B}})
                        }
                    &
                        1
                    &
                        0
                \\
                        \frac{
                            \lambda_{\mathrm{B}}
                            \lambda_{\mathrm{C}}
                        }{
                            (\lambda_{\mathrm{A}} - \lambda_{\mathrm{C}}) (\lambda_{\mathrm{B}} - \lambda_{\mathrm{C}})
                        }
                    &
                        \frac{
                            \lambda_{\mathrm{C}}
                        }{
                            (\lambda_{\mathrm{B}} - \lambda_{\mathrm{C}})
                        }
                    &
                        1
            \end{array}\right]
        .
\end{align}

So the decay of \( \boldsymbol{\alpha}(t) \) from \( t=0 \) until it arrives on the ground at time \( t=\tau \) is
\begin{align}
    \boldsymbol{\alpha}(\tau)
        &=
            \exp\left(
                -
                \boldsymbol{\Lambda}
                \tau
            \right)
            \cdot
            \boldsymbol{\alpha}(0)
        \\
        &=
            \mathbf{P}
            \cdot
            \mathrm{diag}
            \left(
                \exp(-\pmb{\lambda} \tau)
            \right)
            \cdot
            \mathbf{P}^{-1}
            \cdot
            \boldsymbol{\alpha}(0)
        \\
        &=
            \kappa
            \alpha_{\mathrm{Rn}}(0)
            \,
            \mathbf{P}
            \cdot
            \mathrm{diag}
            \left(
                \exp(-\pmb{\lambda} \tau)
            \right)
            \cdot
            \mathbf{P}^{-1}
            \cdot
            \mathbf{e}_{100}
        \\
        &=
            \kappa
            \alpha_{\mathrm{Rn}}(0)
            \,
            \mathbf{P}
            \cdot
            \mathrm{diag}
            \left(
                \mathbf{P}^{-1}
                \cdot
                \mathbf{e}_{100}
            \right)
            \cdot
            \exp(-\pmb{\lambda} \tau)
        \\
        &=
            \kappa
            \alpha_{\mathrm{Rn}}(0)
            \,
            \mathbf{Q}_{100}
            \cdot
            \exp(-\pmb{\lambda} \tau)
        ,
\end{align}
where the final \( \exp \) is taken elementwise over its vector argument, and we have simplified the expression by defining the matrix
\begin{align}
    \mathbf{Q}_{100}
        &\equiv
            \mathbf{P}
            \cdot
            \mathrm{diag}
            \left(
                \mathbf{P}^{-1}
                \cdot
                \mathbf{e}_{100}
            \right)
            \\
            &=
                {\left[\begin{array}{ccc}
                            1
                        &
                            -\frac{
                                \lambda_{\mathrm{B}}
                            }{
                                (\lambda_{\mathrm{A}} - \lambda_{\mathrm{B}})
                            }
                        &
                            \frac{
                                \lambda_{\mathrm{B}}
                                \lambda_{\mathrm{C}}
                            }{
                                (\lambda_{\mathrm{A}} - \lambda_{\mathrm{B}}) (\lambda_{\mathrm{A}} - \lambda_{\mathrm{C}})
                            }
                    \\
                            0
                        &
                            \frac{
                                \lambda_{\mathrm{B}}
                            }{
                                (\lambda_{\mathrm{A}} - \lambda_{\mathrm{B}})
                            }
                        &
                            -\frac{
                                \lambda_{\mathrm{B}}
                                \lambda_{\mathrm{C}}
                            }{
                                (\lambda_{\mathrm{A}} - \lambda_{\mathrm{B}})
                                (\lambda_{\mathrm{B}} - \lambda_{\mathrm{C}})
                            }
                    \\
                            0
                        &
                            0
                        &
                            \frac{
                                \lambda_{\mathrm{B}}
                                \lambda_{\mathrm{C}}
                            }{
                                (\lambda_{\mathrm{A}} - \lambda_{\mathrm{C}}) (\lambda_{\mathrm{B}} - \lambda_{\mathrm{C}})
                            }
                \end{array}\right]}^{\top}\label{eq:Q-transpose}
            .
\end{align}
Note that in \fref{eq:Q-transpose} \( \mathbf{Q}_{100} \) is written as a transpose to save space.


\subsection{Evolution of progeny once on the ground}\label{sec:evolution-on-ground}
Now we consider the second phase where time evolution occurs, that of the progeny activities after they have arrived on the ground.
We will assume that rainfall and all radon progeny densities are uniform over an infinite plane surrounding the detector system.

Let \( \mathbf{a}(t) \) be the areal activity concentrations of the three species on the ground in units of Bq/m\(^{2}\), \( r(t) \) be the instantaneous rainfall rate at ground level in m/s, and \( \boldsymbol{\alpha}(\tau(t)) \) be the progeny activity concentrations in the raindrops at the moment they hit the ground.
Note that \( \boldsymbol{\alpha} \) is written as a function of \( \tau(t) \).
This is because now the time \( t \) is the time according to the detector and it indexes the series of measurements in the rain event, so we are assuming that the apparent rain age \( \tau \) may itself vary over the course of the measurements.

Then \( \mathbf{a}(t) \) follows the differential equation
\begin{align}\label{eq:ground-activity-diff-eqn}
    \dot{\mathbf{a}}(t)
        &=
            r(t)
            \,
            \boldsymbol{\alpha}(\tau(t))
            -
            \boldsymbol{\Lambda}
            \cdot
            \mathbf{a}(t)
        .
\end{align}

Assuming that no radon progeny are initially present, i.e., \( \mathbf{a}(0) = \mathbf{0}_3 \), the solution to this differential equation is
\begin{align}
    \mathbf{a}(t)
        &=
            \int_{t'=0}^{t}
                \exp\left(
                    -\boldsymbol{\Lambda} (t - t')
                \right)
                \cdot
                \boldsymbol{\alpha}(\tau(t'))
                \,
                r(t')
                \,
                dt'
        .
\end{align}

Now we can combine the results of the previous section to express the ground activity concentrations \( \mathbf{a} \) in terms of the physical parameters.
To do this, we will assume that the parameters \( \alpha_{\mathrm{Rn}} \) and \( \tau \) can be indexed by detector time \( t \) --- in other words, that for each detector measurement, \( \tau(t) \) is the apparent age of the rain just arriving at the ground, and \( \alpha_{\mathrm{Rn}}(t) \) is the initial activity concentration of radon in the cloud for the same rain.

Then we obtain
\begin{align}
    \mathbf{a}(t)
        &=
            \int_{t'=0}^{t}
                \mathbf{P}
                \cdot
                \mathrm{diag}
                \left(
                    \exp\left(
                        -\pmb{\lambda}
                        \left(
                            t - t' + \tau(t')
                        \right)
                    \right)
                \right)
                \cdot
                \mathbf{P}^{-1}
                \cdot
                \mathbf{e}_{100}
                \,
                \kappa
                \,
                r(t')
                \,
                \alpha_{\mathrm{Rn}}(t')
                \,
                dt'
        \\
        &=
            \mathbf{Q}
            \cdot
            \int_{t'=0}^{t}
                \exp\left(
                    -\pmb{\lambda}
                    \left(
                        t - t' + \tau(t')
                    \right)
                \right)
                \kappa
                \cdot
                r(t')
                \,
                \alpha_{\mathrm{Rn}}(t')
                \,
                dt'
        .
\end{align}


\subsection{Discretizing the model}\label{sec:discretization}
Since the measurements are taken at discrete time intervals, it will be convenient to discretize the model.
All quantities that are functions of time \( t \) at the ground will be indexed by \( j = 0 \ldots n-1 \) where \( n \) is the number of measurements.
Therefore we get our predicted area concentrations
\begin{align}
    \hat{\mathbf{a}}_j
        &=
            \mathbf{Q}
            \cdot
            \sum_{k=1}^{j}
                \exp\left(
                    -\pmb{\lambda}
                    \left(
                        t_j - t_k + \tau_k
                    \right)
                \right)
                \cdot
                {\left(
                    \kappa
                    r
                    \alpha_{\mathrm{Rn}}
                \right)}_k
                \,
                (
                    t_k - t_{k-1}
                )
        .
\end{align}

We have grouped the rainfall rate~\( r_j \) with \( \kappa \) and the unknown initial cloud activity concentrations \( {\left( \alpha_{\mathrm{Rn}} \right)}_j \).
This was done because the measured rainfall rate for our system, being done at a single point, does not necessarily accurately measure the true rainfall rate in the area around the detector that it is sensitive to, and more work is needed to determine how to define the relevant quantity \( r(t) \) if one desires to accurately infer the activity concentrations in the raindrops and cloud.

So finally we have a model that predicts the ground activity concentrations \( \mathbf{a}_j \) given the parameters \( \tau_j \) and \( {\left( \kappa r \alpha_{\mathrm{Rn}} \right)}_j \), which we will fit to the data.


\subsection{Loss function}\label{sec:loss-function}
Because the physical model generates predicted activity concentrations and estimates the covariances, the statistical portion of the loss function was chosen to be the negative log likelihood of a product of multivariate Gaussians (neglecting constants)
\begin{align}
    {\cal L}_{\mathrm{stat}}
        &=
            \frac{1}{2}
            \sum_{j}
                {(
                    \mathbf{a}_{j}
                    -
                    \hat{\mathbf{a}}_{j}
                )}^{\top}
                \cdot
                \boldsymbol{\Sigma}_j^{-1}
                \cdot
                (
                    \mathbf{a}_{j}
                    -
                    \hat{\mathbf{a}}_{j}
                )
        ,
\end{align}
where we have implicitly removed the first dimension (\isot{Po}{218}) from the model prediction since it is not observed in the gamma-ray data.

Because there are so many free parameters and not all configurations of those parameters are physically plausible, we included regularization terms in the loss.
We took a simple approach of assuming that the quantities, being physical parameters, should be continuous and therefore smooth on some timescale.
The regularization function we chose to impose smoothness is
\begin{align}
    f_{\mathrm{smooth}}(\mathbf{y})
        &=
            \sum_{j=0}^{n-2}
                {\left(
                    \frac{
                        y_{j+1} - y_{j}
                    }{
                        t_{j+1} - t_j
                    }
                \right)}^2
\end{align}
for some data vector \( \mathbf{y} \) and where \( \mathbf{t} \) are measurement times in seconds.

So the total loss function, including both the statistical and regularization portions, was
\begin{align}
    {\cal L}
        &=
            {\cal L}_{\mathrm{stat}}
            +
            \eta_{\alpha_{\mathrm{Rn}} r}
            \,
            f_{\mathrm{smooth}}
            \left(
                \frac{
                    \boldsymbol{\kappa}
                    \,
                    \mathbf{r}
                    \boldsymbol{\alpha}_{\mathrm{Rn}}
                }{
                    {(\kappa r \alpha)}_0
                }
            \right)
            +
            \eta_{\tau}
            \,
            f_{\mathrm{smooth}}
            \left(
                \frac{
                    \boldsymbol{\tau}
                }{
                    \tau_0
                }
            \right)
        ,
\end{align}
where the \( \eta \) coefficients are \( 1 \times 10^{2} \) and \( 1 \times 10^{5} \), which yielded results we judged to be reasonable.
The constants \( {(\kappa r \alpha)}_0 \)\,=\,10\,Bq/m\(^2\)/s and \( \tau_0 \)\,=\,1800\,s were chosen as the relevant dimensional scales of the variables.

The loss function was minimized using JAX~\cite{jax_github_2018} and its optimization library JAXopt~\cite{jaxopt_implicit_diff}, in particular its LBFGS optimizer with parameter bounds.
The bounds chosen were to keep \( \kappa r \alpha_{\mathrm{Rn}} \) non-negative, and to keep \( \tau \) between 1\,minute and 90\,minutes.
Optimization was terminated when the norm of the gradient of the loss function went below 10\(^{-12}\).
On a 2023 MacBook Pro with an Apple M2 Pro chip and 32\,GB of RAM, the optimization took as little as 20\,seconds for a 3-hour rain event, and as long as 20\,minutes for a 24-hour rain event.


\section{Results}\label{sec:results}
\begin{figure*}
    \begin{center}
        \includegraphics[width=0.8\textwidth]{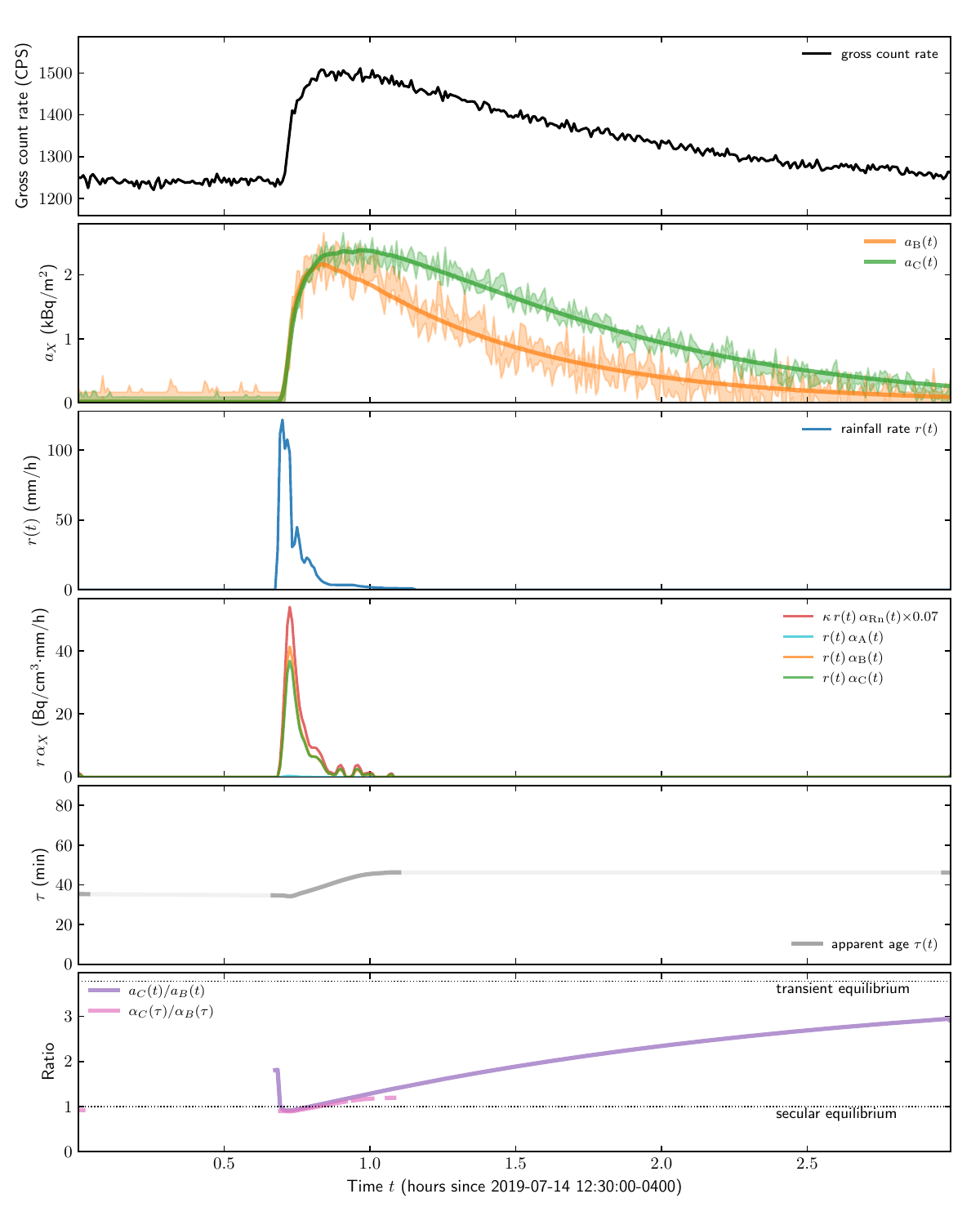}
    \end{center}
    \caption{Analysis of a ``delta-function'' rain event.
    The value of \( \kappa \, r(t) \, \alpha_{\mathrm{Rn}}(t) \) has been scaled by 0.07 for ease of comparison.\label{fig:event-delta}}
\end{figure*}

\begin{figure*}
    \begin{center}
        \includegraphics[width=0.8\textwidth]{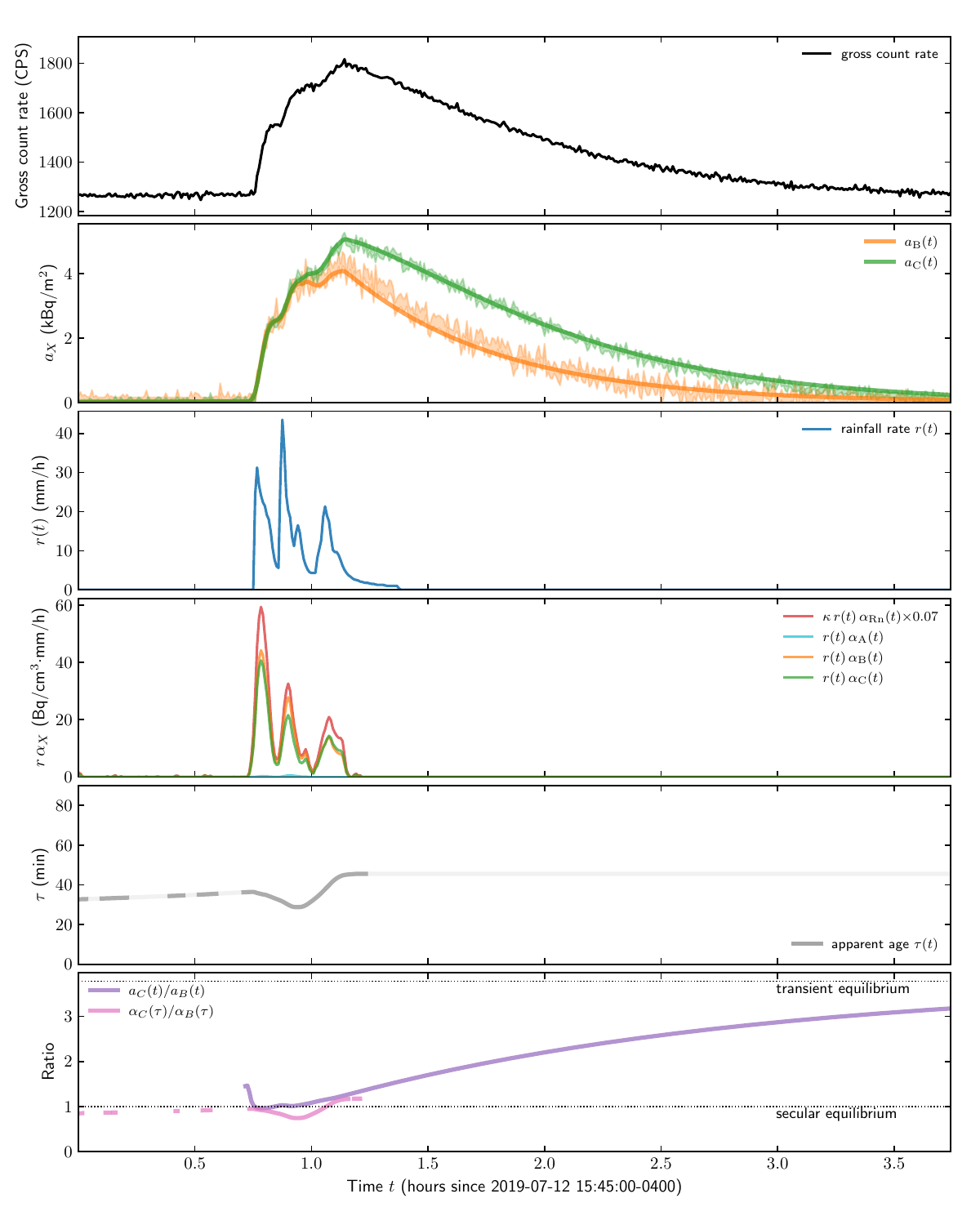}
    \end{center}
    \caption{Analysis of a ``delta-function'' rain event with a more complex structure than~\Fref{fig:event-delta}.
    The value of \( \kappa \, r(t) \, \alpha_{\mathrm{Rn}}(t) \) has been scaled by 0.07 for ease of comparison.\label{fig:event-complex-delta}}
\end{figure*}

The rain events in the dataset can be roughly grouped into approximately three categories:

The first kind are brief events that are less than approximately 1--2 hours long that consist of a brief impulse of intense rainfall and then little to no further rain.
Following ref.~\cite{livesay_rain-induced_2014} we call these ``delta-function'' events.
\Fref{fig:event-delta} shows one such delta-function event where there is one strong intense episode of rain.
Some of these events contain multiple closely spaced, intense impulses of rain rather than a single impulse.
\Fref{fig:event-complex-delta} shows one such event, which has three strong rain impulses, and the total rainfall lasts no longer than 1.5~hours.

\begin{figure*}
    \begin{center}
        \includegraphics[width=0.8\textwidth]{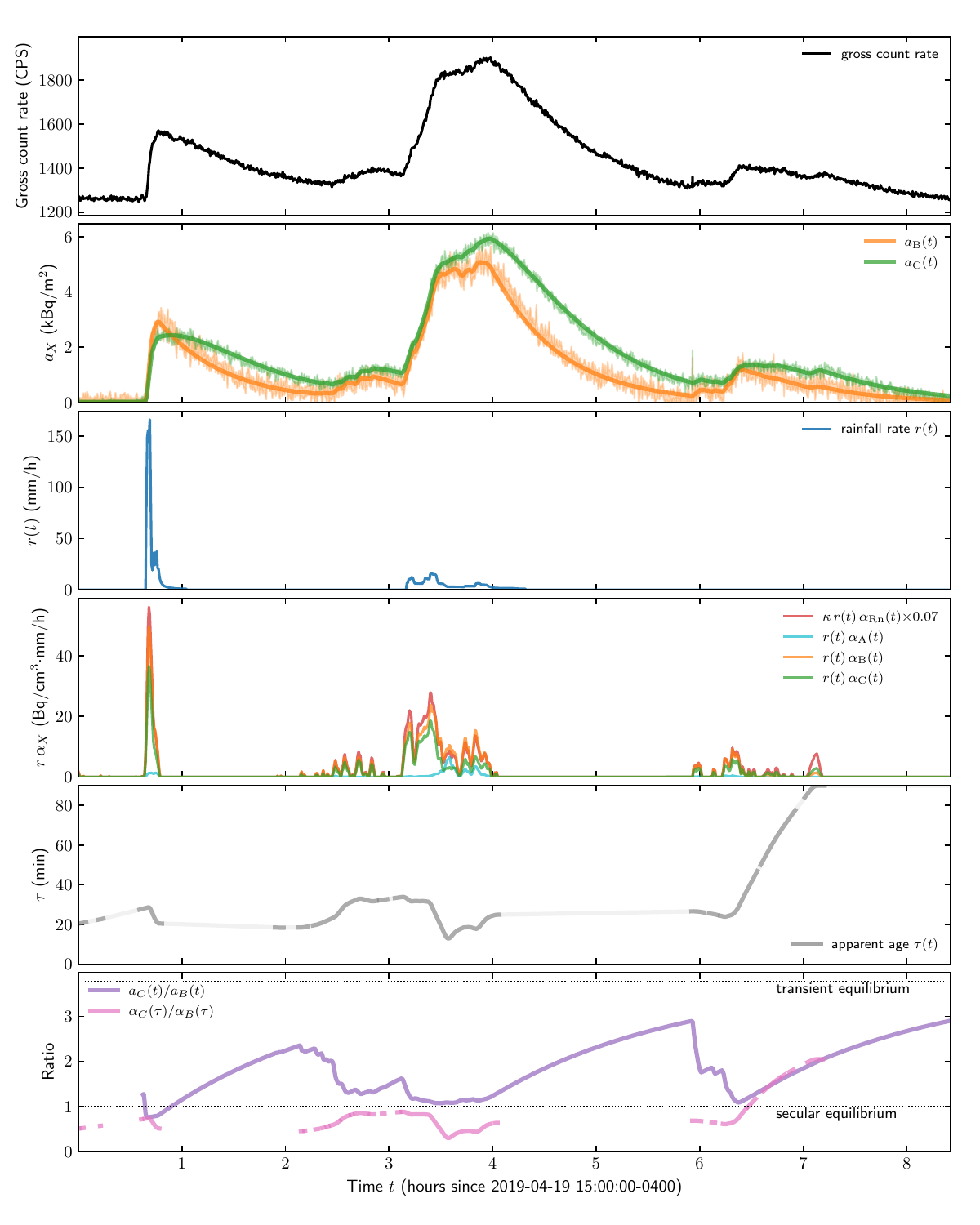}
    \end{center}
    \caption{Analysis of a rain event with multiple brief episodes.
    The value of \( \kappa \, r(t) \, \alpha_{\mathrm{Rn}}(t) \) has been scaled by 0.07 for ease of comparison.\label{fig:event-multi-delta}}
\end{figure*}

The second category consists of multiple delta-function type events that are separated in time but whose decay progeny signals overlap.
\Fref{fig:event-multi-delta} shows an event made up of about three short events, and it lasts several hours in total.
These events are likely part of the same weather system.

\begin{figure*}
    \begin{center}
        \includegraphics[width=0.8\textwidth]{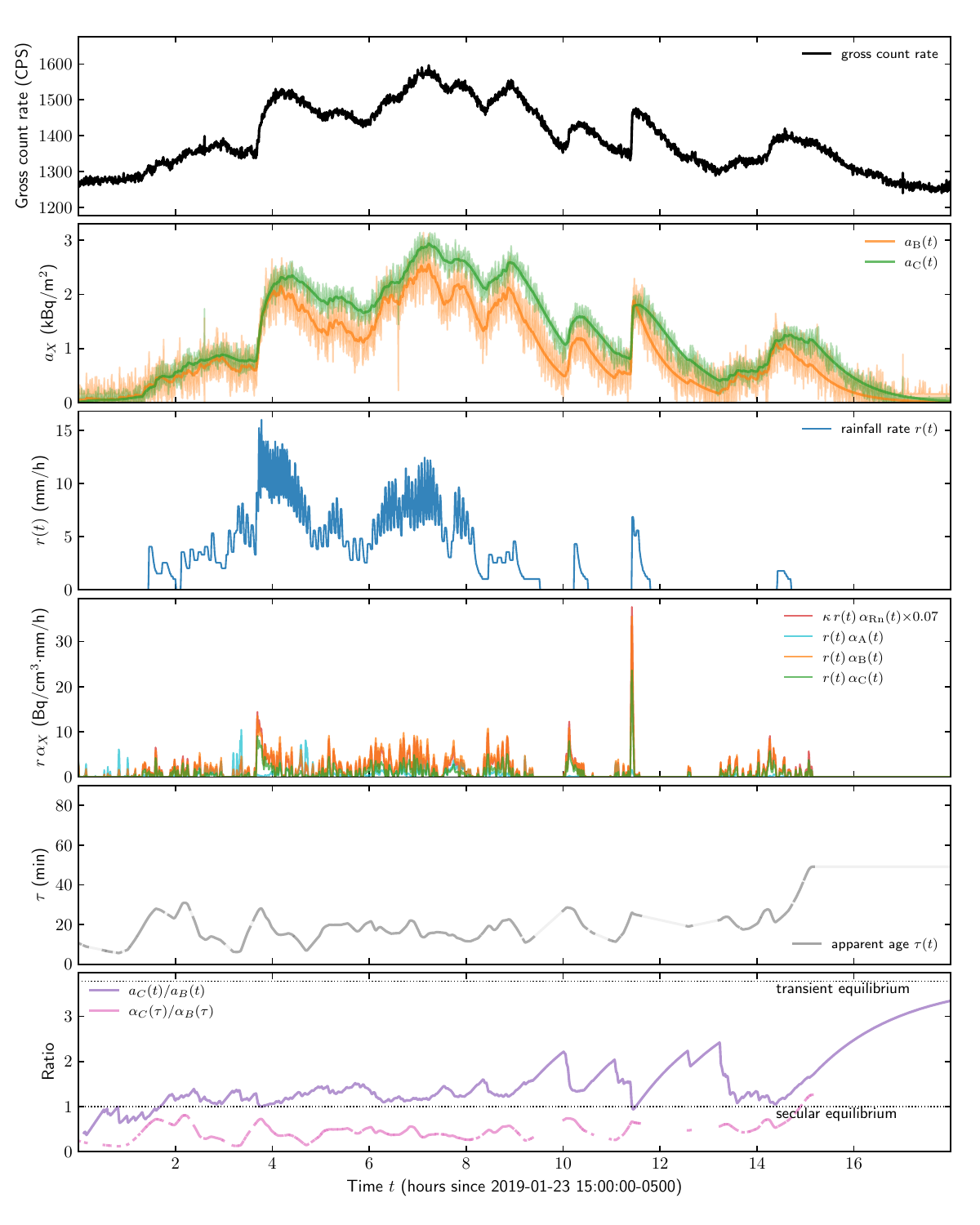}
    \end{center}
    \caption{Analysis of a long complex rain event.
    The value of \( \kappa \, r(t) \, \alpha_{\mathrm{Rn}}(t) \) has been scaled by 0.07 for ease of comparison.\label{fig:event-long-complex}}
\end{figure*}

The final category are long complex events, where the rainfall is too continuous to be divided into a series of short events.
These events generally consist of lower rainfall rates spread out over several hours to over 24~hours.
\Fref{fig:event-long-complex} shows an example where a low rate of rainfall lasts about 12~hours.
Events from this category typically occur in the winter at this location, while the other event types are more common in the summer.
This phenomenon of shorter, more intense rainfall events in the summer and longer, less intense rainfall events in the winter is observed in other temperate locations, such as Japan~\cite{hayakawa_radon-concentration--cloud_1985,minato_simple_2007,moriizumi_214bi214pb_2015}.

For each of the events shown in \Fref{fig:event-delta}--\ref{fig:event-long-complex}, a variety of measurements and model parameters are shown.
The top plot shows the gross count rate of the gamma-ray spectrum, which always rises during the rain event.
The second plot shows the activity concentrations on the ground, \( a_{\mathrm{B}} \) and \( a_{\mathrm{C}} \), which are the source terms needed for the production of synthetic data, the main object of this work.
The filled regions of the same color show the estimated values plus-or-minus one standard deviation, where these quantities were found using the methods described in \Fref{sec:estimating-densities}.

The third plot shows the rainfall rate measured by the weather station, and the plot below that shows the rainfall rate times activity concentrations.
Sometimes, as in \Fref{fig:event-delta}, the rainfall rate closely tracks the latter quantities and therefore could plausibly be divided to obtain the bare activity concentrations \( \alpha_{X} \).
Doing so would result in activity concentrations \( \alpha_{\mathrm{B}} \) and \( \alpha_{\mathrm{C}} \) on the order of 1\,Bq/cm\(^3\), which is in agreement with published values (e.g., refs.~\cite{damon_natural_1954,minato_estimate_1983,fujinami_influence_1985,nishikawa_automatic_1986,paatero_wet_2000,cortes_automated_2001,horng_situ_2004,takeyasu_concentrations_2006}).
However, we left the rainfall rate undivided since the gamma-ray data frequently indicated that rainfall was present when none was measured by the weather station (e.g., \Fref{fig:event-multi-delta}), which can occur because the gamma-ray detector is sensitive to the surrounding area out to hundreds of meters, while the weather station measures rainfall over an area of only a few~cm\(^{2}\).

The final two plots are the apparent radiometric age \( \tau \) and the ratios of the activity concentrations in rainwater in the air (\( \alpha_{\mathrm{C}} / \alpha_{\mathrm{B}} \)) and on the ground (\( a_{\mathrm{C}} / a_{\mathrm{B}} \)).
The age results are plausible, with most having ages that are at least several minutes to tens of minutes~\cite{horng_situ_2004,greenfield_determination_2008,moriizumi_214bi214pb_2015}.
Because of the assumptions of our model, the age directly determines the activity concentration in the rain when it reaches the ground (\( \alpha_{\mathrm{C}} / \alpha_{\mathrm{B}} \)), and those values fall between \( \approx \)\,0 and \( \approx \)\,2, which is typical of other observations~\cite{finck_situ_1980,rangarajan_observations_1985,horng_situ_2004,takeyasu_concentrations_2006,moriizumi_214bi214pb_2015}.

The ratio of activity concentrations on the ground is brought down to \( \approx \)\,1 when rain is falling, and when there is no rain it trends toward the transient equilibrium value \( \lambda_{\mathrm{C}} / ( \lambda_{\mathrm{C}} - \lambda_{\mathrm{B}} ) \)\,=\,3.78, which is what is expected from~\fref{eq:ground-activity-diff-eqn} when \( \boldsymbol{\alpha} \)\,=\,\(\mathbf{0}\) and \( t \)\,\(\rightarrow\)\,\(\infty \)~\cite{finck_situ_1980}.
The overall magnitude of the deposited activity concentration is on the order of 1\,kBq/m\(^2\) for both species, which agrees with other observations~\cite{finck_situ_1980,hirouchi_estimation_2014}.

In general, the physical model is good fit to the data, and the physical parameters are in reasonable ranges, with the caveat that we cannot fully evaluate the values of \( \alpha_{\mathrm{Rn}} \) without unfolding by the rainfall rate and the air-to-raindrop volume factor \( \kappa \).


\section{Discussion}\label{sec:discussion}
We developed a quantitative full-spectrum model of the background of a stationary gamma-ray detector situated outdoors.
This model pays special attention to the radon progeny \isot{Pb}{214} and \isot{Bi}{214}, and it estimates their activity concentration on the ground around the detector.
A physical model that accounts for the radioactive decay of the radon progeny is used to fit the results and estimate the age of the radon progeny and their concentration in the raindrops.
The results we obtain are broadly consistent with other observations of the same phenomena.

The methods described in~\Fref{sec:modeling-detector} and \Fref{sec:modeling-radon-progeny} and the results from~\Fref{sec:results} were accurate and quantitative enough to use as the source terms in synthetic urban search datasets, which was the primary purpose of this research.
We identified a number of avenues for further research that could improve the model further and describe them here.

The validation of the model consisted of checking the activity concentrations against ranges from the literature, but it could be improved by taking a ground truth in the specific location.
This could be done by sampling the soil in a grid around the detector and measuring the specific radioactivity of the terrestrial isotopes.
The radon progeny would be more difficult to properly ground truth because of their short half-lives.
However, \isot{Be}{7} could be used as a stand-in for those and be assayed along with the terrestrial isotopes.
It may also be useful to calibrate using check sources placed on the surface of the ground at various distances, for example, using a \isot{Eu}{152} source with many photopeaks spread across the entire spectrum.
The ground truth method would need to understand the systematics due to surface roughness, which affect \isot{Pb}{214} more because of its lower energy emission.

Another improvement is to obtain estimates of the uncertainties of the model parameters and outputs.
An attempt was made here to calculate the Hessian of the loss function using JAX, which would allow the calculation of the Fisher information matrix and its inverse, which would in turn allow us to estimate the covariance matrix of the model parameters assuming they meet the Cram\'{e}r-Rao lower bound~\cite{kendalls_advanced_statistics}.
JAX also allows the easy calculation of the various Jacobians needed to find the uncertainties of quantities derived from the model, such as \( \alpha_{\mathrm{B}} \) and \( \alpha_{\mathrm{C}} \).
The problem with this method is that, with thousands of parameters in some cases, the calculation of the Hessian matrix using JAX required more than 120\,GB of memory, which exceeded what was available on the machine.
Some other, less resource-intensive approach (or a machine with a large amount of RAM) would be needed to estimate the uncertainties using this approach.

A more accurate and robust method for measuring the rainfall rate \( r(t) \) should also be explored, as discussed earlier.
In the model, \( r(t) \) must reflect the rainfall rate of any rain that produces a signal in the NaI(Tl) detector, and measuring \( r(t) \) using a single rain gauge can evidently miss nearby precipitation that nevertheless registers with the detector (e.g., \Fref{fig:event-multi-delta}).

There are some other sensor technologies that could be used to obtain better measurements of rainfall in the surroundings.
One class of such sensors are ``passive'' sensors that, like rain gauges, ``listen to'' the environment without emitting their own signals.
Acoustic sensors are one example, although they are most sensitive to rain that directly impacts them~\cite{trono_rainfall_2012,avanzato_innovative_2020}.
Still, an array of these sensors could more easily be deployed in an area around the detector than an array of rain gauges.
Likewise, seismic sensors have been shown to detect signatures from rainfall~\cite{dean_seismic_2017}, with 90\% of the measured seismic power coming from within 5--25\,m of each sensor~\cite{bakker_seismic_2022}, making them sensitive to a greater area than acoustic sensors.
An array of such sensors embedded in the ground around the detector would also be needed.
Audio and acoustic recordings have been shown to be able to detect and estimate the rate of rainfall and would be sensitive to an even wider area~\cite{bedoya_automatic_2017,wang_rainfall_2022}.
Another possibility is the use of video, which works by measuring the optical streaks left by raindrops and estimating their size and velocity, and inferring other properties~\cite{allamano_toward_2015,dong_measurements_2017,wang_novel_2022,yan_review_2023}.
However, these methods are generally experimental and have high uncertainties.

``Active'' sensors that emit signals and look for returns are also possible options.
Traditional radar methods generally have too poor a spatial and temporal resolution for this purpose~\cite{berne_temporal_2004}.
However, lidar can work on the scale of a few hundred meters~\cite{lewandowski_lidar-based_2009}.
Only certain types and configurations of lidar make these measurements feasible; the increasingly common and affordable automotive-style lidars are largely insensitive to precipitation~\cite{filgueira_quantifying_2017,goodin_predicting_2019}.

Looking to the future, an emerging generation of networked multi-sensor gamma-ray systems for radiological and nuclear security are being tested and deployed in different areas~\cite{archer_radiation_2019_minos3,cooper_networked_2023}.
These systems will provide a wealth of data that would also be useful for dual-purpose studies in atmospheric science, health physics, and other disciplines.
Their combination of large NaI(Tl) detectors with cameras, lidar, and weather stations spread out at many points in a city means that not only could radon progeny be studied at high frequencies (measurements faster than 1/minute) and high spatial resolution (\(\sim\)100\,m), but such measurements could also be correlated over larger areas (\(\sim\)10\,km), which would allow for the observation of the variability of atmospheric parameters on the scale of an entire cloud or weather front.


\section{Conclusions/Summary}\label{sec:conclusions}
We have presented an analysis of a set of measurements of a NaI(Tl) detector located outdoors during a series of rainfall events.
To quantitatively analyze the spectra, we developed full-spectrum models of each of the background components, including the radon progeny \isot{Pb}{214} and \isot{Bi}{214}.
We also required that the activity concentrations of the radon progeny follow a physical model of radioactive decay, and in doing so were able to estimate some quantities relevant to atmospheric studies.
Aside from being useful for producing radon progeny source terms that add more realism to synthetic datasets for search applications, this method could also be applied in other areas, such as atmospheric science or agricultural studies, to estimate quantities of interest with both greater accuracy and higher temporal frequency.


\section{Acknowledgments}\label{sec:acknowledgments}
This research was supported by the U.S. National Nuclear Security Administration (NNSA) Office of Defense Nuclear Nonproliferation Research and Development within the U.S. Department of Energy, and performed by Lawrence Berkeley National Laboratory (LBNL) under Contract DE-AC02-05CH11231, and by Oak Ridge National Laboratory, managed by UT-Battelle, LLC under Contract DE-AC05-00OR22725.

The code used in the analysis for this manuscript was written in Python.
In addition to JAX~\cite{jax_github_2018} and JAXopt~\cite{jaxopt_implicit_diff} for optimization, it also relied heavily on NumPy~\cite{harris_numpy_2020} and SciPy~\cite{scipy_2020} for computation, Matplotlib~\cite{hunter_matplotlib_2007} for plotting, and Becquerel~\cite{bandstra_becquerel_2021} for nuclear data and spectrum manipulations.


\printcredits



\bibliographystyle{elsarticle-num-names}
\bibliography{rain_modeling_paper}


\end{document}